\documentstyle[12pt,aaspp4,epsf]{article}
\lefthead{Nagar \& Wilson}
\righthead{Radio-extended Seyferts}
\begin{document}


\title{The Relative Orientation of Nuclear Accretion and Galaxy Stellar 
Disks in Seyfert Galaxies}
\author{Neil M. Nagar and Andrew S. Wilson}
\affil{Department of Astronomy, University of Maryland, College Park, MD
20742; neil@astro.umd.edu, wilson@astro.umd.edu}
\received{September 14, 1998}
\accepted{November 17, 1998}
\begin{center}
\textbf{To appear in ApJ, Vol 516 \#1, May 1, 1999}
\end{center}

\begin{abstract}

We use the difference ($\delta$) between the position angles of 
the nuclear radio emission and the host
galaxy major axis to investigate the distribution of the angle ($\beta$) 
between the axes of the nuclear accretion disk and the host galaxy disk in
Seyfert galaxies.
We provide a critical appraisal of the quality of all measurements, and find
that the data are limited by observational uncertainties and biases,
such as the well known deficiency of Seyfert galaxies of high inclination.
There is weak evidence that the distribution of $\delta$ for Seyfert~2 galaxies
may be different (at the 90\% confidence level) from a uniform distribution,
while the Seyfert~1 $\delta$ distribution is not significantly different from
a uniform distribution or from the Seyfert~2 $\delta$ distribution.
The cause of the possible non-uniformity in the distribution of $\delta$ for
Seyfert~2 galaxies is discussed.
Seyfert nuclei in late-type spiral galaxies may favor large values of $\delta$ 
(at the $\sim$96\% confidence level), while those in 
early-type galaxies show a more or less random distribution of $\delta$. 
This may imply that the nuclear accretion disk in non-interacting late-type 
spirals tends to align with the 
stellar disk, while that in early-type galaxies is more randomly oriented,
perhaps as a result of accretion following a galaxy merger.

We point out that biases in the distribution of inclination translate to 
biased estimates of $\beta$ in the context of the unified scheme. When this
effect is taken into account, 
the distributions of $\beta$ for all Seyferts together, and of Seyfert~1's and
2's separately, agree with the hypothesis that the radio jets
are randomly oriented with respect to the galaxy disk. 
The data are consistent with the expectations of the unified scheme, but
do not demand it.
\end{abstract}


\section{Introduction}
The nuclear radio sources in Seyfert galaxies often show a ``linear'' 
(i.e. double, triple or jet-like) structure on the tens of pc to kpc
scale. Such sources are believed to result from collimated ejection of 
radio-emitting plasma by the nucleus (Wilson \& Willis 1980), presumably
along the rotation axis of the nuclear accretion disk (e.g. Pringle 1993,
Blandford 1993). The observed orientations of the linear radio sources
can then be used to probe the distribution of the angle ($\beta$; we use
the nomenclature of Clarke, Kinney \& Pringle 1998, hereafter C98) between
the nuclear accretion disk and the stellar disk of the host galaxy. This 
distribution may provide clues to the origin of accretion disks in active
spirals. 

The distribution of the angle, $\delta$, between the direction on the sky of
the radio sources and the apparent major axis of the galaxy disk (see Fig. 1)
has been investigated by Ulvestad \& Wilson (1984b, hereafter Paper VI), 
Schmitt et al. (1997, hereafter S97) and C98.
In Paper VI, a weak trend was found for the distribution of $\delta$ to avoid
values near 0{\arcdeg} or 90{\arcdeg}, but this trend was not statistically
significant given the small size of the sample.
S97 found that Seyfert~1's are less likely to have extended radio structures 
along the host galaxy major axis (i.e. values of $\delta$ near 0{\arcdeg})
while Seyfert~2's have these structures distributed in most directions.
They also concluded that both Seyfert types seem to avoid close alignment 
between radio and galaxy disk axes (i.e. values of $\delta$ near 90{\arcdeg}).
S97 interpreted their results in terms of a model in which Seyfert~1's have
the axis of their accretion disk aligned preferentially along the host
galaxy disk axis while the accretion disk axis in Seyfert~2's can
assume any angle relative to the host galaxy plane, with the
exception of those angles which result in an observed $\delta$
$>$~70{\arcdeg}.
This is consistent with an earlier model proposed by Schmitt \& Kinney (1996), 
in which the axis of the accretion disk in Seyfert~1's  may be aligned 
preferentially along the rotation axis of the host galaxy. 
The analysis of Schmitt \& Kinney (1996) was based on the spatial extent of 
line emission in archival HST images of Seyferts, finding a much greater 
extent for Seyfert~2's than 1's. However, the galaxies selected for these 
early (pre-COSTAR) observations by HST were preferentially Seyfert~2's known
to have extended emission lines in ground-based observations (Wilson 1997),
so the very large difference in sizes found should be treated with 
caution\footnote{More explicitly, most of the pre-COSTAR HST emission-line
images of Seyfert galaxies used by Schmitt \& Kinney (1996)
were taken in three programs - the FOS GTO program, the FOC GTO program
and GO program 3724. Two galaxies were observed in other 
programs. The FOS team selected a mixture of Seyfert~1's and 2's 
without an obvious bias towards galaxies known to have extended 
emission-line regions in ground-based observations. On the other hand, the
galaxies imaged in the FOC GTO and GO~3724 were preferentially
selected to be Seyfert~2's with extended emission-line regions in 
ground-based observations. Seyfert~2's were preferentially observed 
in order to avoid confusion by the PSF of the strong compact
nuclear source of Seyfert~1's. Galaxies with extended emission lines
in ground-based observations were selected to ensure an ``informative''
image with HST. Thus the Seyfert~1's
and Seyfert~2's studied by Schmitt \& Kinney (1996) were selected
according to different criteria, and the sample is not well suited
to investigating differences in their extended emission-line regions.
As they noted, a sample selected by an isotropic property would be 
required for a conclusive result.}.

\begin{figure}[!ht]
\figurenum{1}
\plotfiddle{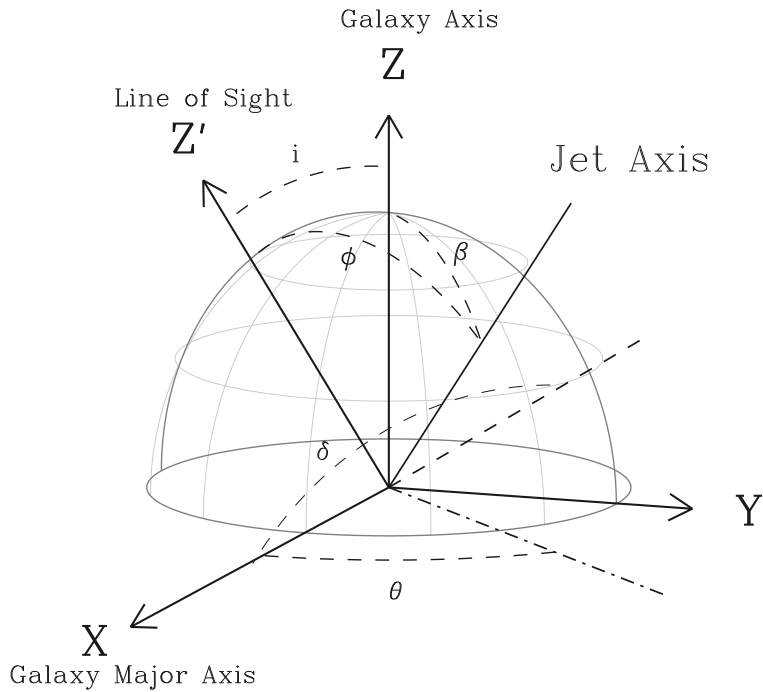}{5in}{0}{92}{92}{-290}{-230}
\vspace{-1.3in}
\caption
{\small{ 
Diagram of the geometry, following the nomenclature of C98.
The galaxy stellar disk lies in the X, Y plane and the apparent major axis of
the galaxy disk is along the X axis. The observer is along the Z$^\prime$ axis,
an angle of $i$ from the Z axis. The axis of the radio jet, its projection
onto the plane of the sky (heavy dashed line) and its projection onto the
galaxy disk (dot-dashed line) are shown. The angles $\beta$ (between the 
jet and galaxy disk axis), $\delta$ (between the projection of the jet
onto the plane of the sky and the apparent major axis of the galaxy disk),
$\theta$ (between the projection of the jet onto the galaxy disk and the
apparent major axis of the disk) and $\phi$ (between the jet and the line
of sight) are indicated. Positive values of $\phi$ result when the jet's 
projection on the sky falls on the far side ($Y~>~0$) of the disk, 
and negative values of $\phi$ when the jet's projection on the sky falls
on the near side.
}}
\end{figure}

C98 have shown how measurements of both the galaxy inclination ($i$) and 
$\delta$ can be used to obtain better estimates of the distribution of
$\beta$ than if only $\delta$ is used. In a comprehensive analysis, they found
that the distribution of $\beta$ for the whole sample (Seyfert~1's and 2's
taken together) is consistent with a uniform distribution in $\cos~\beta$,
i.e. the accretion disks are randomly oriented w.r.t. the galaxy stellar disk.
They also investigated the distributions of $\beta$ for Seyfert~1's and 2's 
separately, assuming the validity of the unified scheme. In this scheme
(see, e.g., Antonucci 1993), 
Seyfert~1's are supposed to be galaxies in which the angle, $\phi$, 
between the line of sight and accretion disk axis is less than some canonical
value, $\phi_c$, while Seyfert~2's have values of $\phi$ greater than $\phi_c$.
The value of $\phi_c$ is believed to be about 30{\arcdeg} based on the 
relative number of Seyfert~1 and Seyfert~2 galaxies (e.g. \cite{ostsha88}).
Assuming that the value of $\phi$ is the only difference between Seyfert~1's
and 2's, C98 found that the distributions of $\beta$ for Seyfert~1's and 2's
are different at the 96\% confidence level (i.e. approximately 2$\sigma$), with
Seyfert~1's favoring low values and Seyfert~2's high values of 
$\beta$. They concluded that $\phi_c$ may not be a universal constant or the
$\beta$ distributions of Seyfert~1's and 2's are significantly different. 
In the present paper (Section 3.2), however, we argue that a bias in
the distribution of $i$ for Seyfert galaxies can account for the apparent
difference in the $\beta$ distributions of Seyfert~1's and 2's, and that
there is no evidence that the $\beta$ distributions of the two Seyfert 
types differ from each other.

Any observational study of the relative orientation of the radio and galaxy 
disk axes in Seyfert galaxies is limited by small numbers and large 
observational uncertainties and biases.  In this paper, we reevaluate the
distribution of $\beta$ following the technique devised by C98. We 
include new radio maps of Seyferts from a recent survey by Nagar et al. 
(1998, hereafter Paper VIII), and provide a critical appraisal of all
measurements of radio and galaxy major axes and galaxy inclinations.
Section 2 describes the samples and data used in this paper, and provides a
detailed discussion of the uncertainties in the measurements of the
radio axis and host galaxy major axis. 
Section 3 presents the results and Section 4 provides a brief discussion.
Distance dependent quantities are calculated using H$_0$~= 50 km s$^{-1}$
Mpc$^{-1}$.

\section{Samples \& Measurements}
\subsection{The Samples}
The analysis used in this paper requires that we know the P.A.'s of the nuclear
radio sources and host galaxy major axes, and the inclination of the host 
galaxy disk. We determine these parameters for two samples.
The first sample comprises all Seyferts with extended nuclear radio
structure in the distance- and magnitude-limited sample of early-type Seyfert 
galaxies of Paper VIII. This sample consists of all Seyferts known as of 1991
with total magnitude m$_V$~$\leq$14.5, recessional
velocity cz~$<$~7,000 km s$^{-1}$, morphological type E, S0 or S0/a,
and declination north of $\delta$=$-$41{\arcdeg}. It numbers
43 Seyfert galaxies (14 Seyfert~1's, 2 Seyfert~1.9's and 27 
Seyfert~2.0's\footnote{We use `Seyfert~1' to denote Seyfert~1.0 through
Seyfert~1.5, and `Seyfert~2.0' for galaxies without broad wings on any 
permitted lines. Seyfert~1.8's and 1.9's are treated separately, as
these could represent Seyfert 1's reddened by dust in the plane of the
galaxy disk outside of the putative circumnuclear torus
(see, e.g., Lawrence \& Elvis 1982, Keel 1980 and Maiolino \& Rieke 1995).
}). 
All these objects have been observed with the VLA at both 3.6~cm and 20~cm with
resolutions of $\sim$0{\farcs}3 and $\sim$1{\farcs}5, respectively.
To create the second sample used in this paper - the ``radio-extended Seyfert 
sample'' - we added to the early-type Seyfert sample all other Seyferts in
the literature for which high-resolution ($\lesssim$~2{\arcsec}) radio imaging 
shows extended nuclear radio structure.
The radio-extended Seyfert sample consists of 75 Seyferts
(26 Seyfert~1's, 9 Seyfert~1.8's and 1.9's, and 40 Seyfert~2.0's).
This sample includes the 46 Seyferts in the list compiled by S97.
Sixty six percent of the 75 galaxies belong to either
or both of the distance- and magnitude-limited early-type Seyfert sample of 
Paper VIII and the distance-limited Seyfert sample of Ulvestad \& Wilson (1989,
hereafter Paper VII); most of the remaining Seyferts are Markarian galaxies
from Ulvestad \& Wilson (1984a, hereafter Paper V).
We emphasize that such a sample must contain several selection effects, 
and that a future sample selected by a more isotropic property would be
better suited for such a study.

\vspace{-0.1in}

\subsection{Difficulties in determining the radio axis}
The most reliable radio axes come from sources classified `L'
(i.e. double, triple or jet-like sources straddling the optical
nucleus). Even here, observations have shown that the 
apparent axis of the radio ejecta can change significantly with
increasing distance from the nucleus. For example, in 
NGC~1068, the ejecta on a scale of $\simeq$10~pc lie in 
P.A.$\simeq$7{\arcdeg} (e.g. Gallimore, Baum and O'Dea 1996),
but change to $\simeq$30{\arcdeg} at distances of 30$-$400~pc
(e.g. \cite{wilulv87}).
In another case, VLBA imaging of the nucleus of NGC~4151 reveals what may
be a `bend' in the jet, from P.A. 20{\arcdeg} at the 
$\simeq$40~pc scale to P.A. 75{\arcdeg} at the $\simeq$280~pc scale
(\cite{ulvet98}), though an interpretation of
the inner radio sources as part of a disk or torus cannot be ruled out.
For typical distances to Seyfert galaxies and the resolution of the VLA, 
only the hundred~pc$-$kpc scale structure can be resolved, with
consequent uncertainty in the true nuclear axis.
The situation is worse for class `S' (slightly resolved) sources, where it
may be unclear whether the extended radio emission represents ejecta from the
nucleus or disk radio emission or a mixture of both.
The percentage of resolved radio sources which are  of
class `L' or `(L)' (the parentheses indicate uncertain classification)
in the sample of early-type Seyferts, the sample of all radio-extended
Seyferts, and the sample used by S97 are 53\%, 52\% and 67\%, respectively. 
The lack of `L' or `(L)' class sources among Seyfert~1's
is particularly acute; in the early-type sample, the sample of all 
radio-extended Seyferts, and the sample of S97, there are only 5 (36\% of
extended radio sources in Seyfert~1's), 
11 (42\%) and 8 (53\%) such sources, respectively.  We have attempted to
provide a measure of the reliability of each radio source P.A. by means of 
a `quality flag'. 
`L' class radio sources are assigned quality flag `a', `S' class radio
sources with high S/N ratio are assigned quality flag `b', and `S' class
radio sources with low S/N ratio, or doubtful extension are assigned quality
flag `c' (see Table 4 of Paper VIII).

We note that, while the three Seyfert~1's with
the highest recessional velocities in the radio-extended Seyfert sample
all have very small values ($\leq$10{\arcdeg})
of $\delta$, their  radio structures are very likely to be nuclear in origin
as they all show well defined `L' class radio structure at high resolutions - 
1$-$5 milli-arcsec (1$-$6~pc, Mrk~231, \cite{ulvet97}),  
$\sim$0{\farcs}2 ($\sim$230~pc, UGC~5101, \cite{sopale91}), and
0{\farcs}25 (350~pc, Mrk~926, \cite{wil96}).  

\vspace{-0.1in}

\subsection{Difficulties in determining the galaxy major axis}
The P.A. of the major axis of the galaxy disk is also often uncertain.
Seyfert galaxy disks may contain bars and oval distortions, leading
to incorrect axes if these are interpreted as projections of inclined,
circular disks (see e.g. \cite{davies73} for NGC~4151 and Baldwin,
Wilson \& Whittle 1987
for NGC~1068). Nearby companion galaxies may also distort the outer
isophotes of Seyfert galaxies.

The major axis of the galaxy should ideally be determined kinematically.
Kinematic major axes are available from optical emission-line
mapping of the gas rotating in the galaxy disk in four Seyferts - 
NGC~2110 (multiple long-slit spectra; \cite{wb85}), 
NGC~3081 (imaging Fabry-Perot; \cite{bupu98}),
NGC~5643 (imaging Fabry-Perot; \cite{moret85}) and
NGC~5728 (imaging Fabry-Perot; \cite{schet88}) - and from
stellar kinematics in one Seyfert - NGC~3516 (bidimensional spectroscopy;
\cite{arret97}). 
HI 21~cm kinematic maps exist for 11 Seyferts (NGC~1068, NGC~1365, 
NGC~3227, NGC~3627, NGC~4051, NGC~4151, NGC~4258, NGC~5033, NGC~6300,
NGC~6814 and Mrk~348).
We compared the HI kinematic major axis P.A. for these 
galaxies with the photometric major axis P.A. listed in the
``Third Reference Catalogue of Bright Galaxies'' 
(de Vaucouleurs et al. 1991, hereafter RC3), 
and found that the two values of the P.A. were in good agreement for
inclined galaxies (for which the log of the ratio of major to
minor axes as defined in RC3 - log~R$_{25}$ - was~$>$~0.3), 
while the difference between the two P.A.'s was larger (up to 
45{\arcdeg} for NGC~1068) for more face-on galaxies
(log~R$_{25}$~$<$~0.3).
For this reason, we assign quality flag `a' to all
kinematic measurements of major axes, and to photometric measurements
when the galaxy is highly inclined (log R$_{25}~>$~0.3). Photometric major
axes for galaxies with 0.04~$\leq$~log~R$_{25}$~$<$~0.3 are assigned quality
flag `b'. Galaxies for which no P.A. is listed in RC3, the 
``Uppsala General Catalogue of Galaxies'' (Nilson 1973, hereafter UGC) 
or the ``The ESO/Uppsala Survey of the ESO(B) Atlas'' (Lauberts 1982,
hereafter ESO), and for which
a major axis has been obtained from other sources, are assigned quality flag
`c'. Quality flag `c' also includes objects with disturbed or peculiar 
morphology. Quality flag `d' is given to galaxies with log~R$_{25}~<$~0.04 
or highly disturbed morphology or which are strongly interacting; such
objects are omitted from our study. These flags are summarized in Table~4 
of Paper VIII.

\subsection{The Data}
The P.A.'s of the radio and galaxy major axes (P.A.$_{Radio}$ and
P.A.$_{Galaxy}$) for the early-type Seyfert sample are listed in Table~3 of 
Paper VIII. We also use the P.A.'s of the major axes of the galaxies in the
green continuum at a surface brightness of 2~$\times$~10$^{-18}$
erg cm$^{-2}$ s$^{-1}$ \AA$^{-1}$ (arcsec)$^{-2}$ (P.A.$_{Green}$, from 
CCD imaging by Mulchaey, Wilson \& Tsvetanov 1996, hereafter MWZ)
as a second measure of the host galaxy major axes. This surface brightness
is higher than used for the first measure of host galaxy major axis and
is thus more likely to be affected by bars or oval distortions. However, 
the P.A.'s
at this brightness represent a more uniform data set and are available for all
galaxies in the early-type sample. A histogram of the difference 
P.A.$_{Green}$ $-$ P.A.$_{Galaxy}$ (Figure~2) shows a strong trend for 
alignment between these two axes, but also a number of outliers.
Brief notes on the radio and galaxy P.A.'s for most objects in
the early-type sample can be found in Section 4 of Paper VIII.

\begin{figure}[!ht]
\figurenum{2}
\plottwo{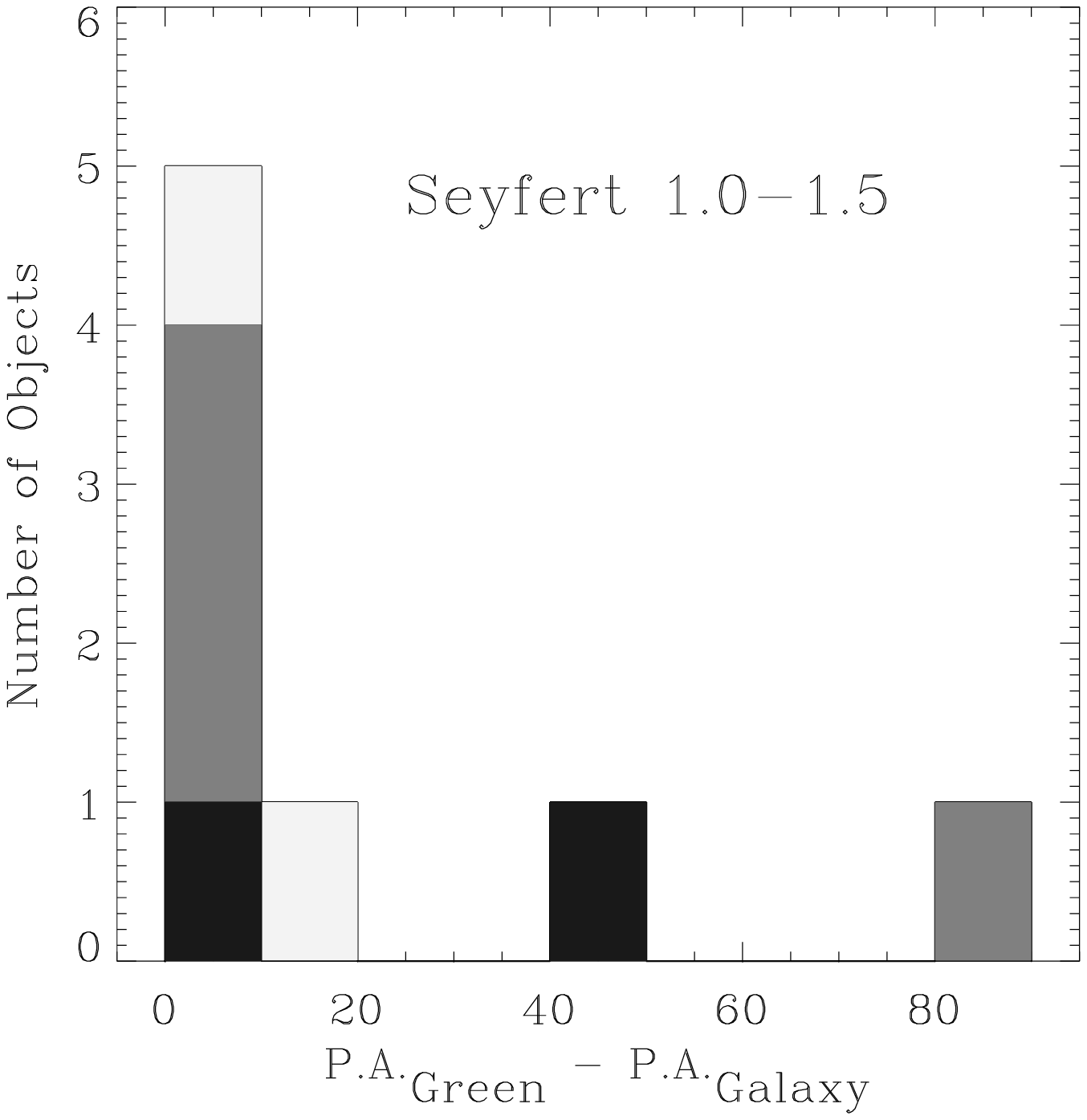}{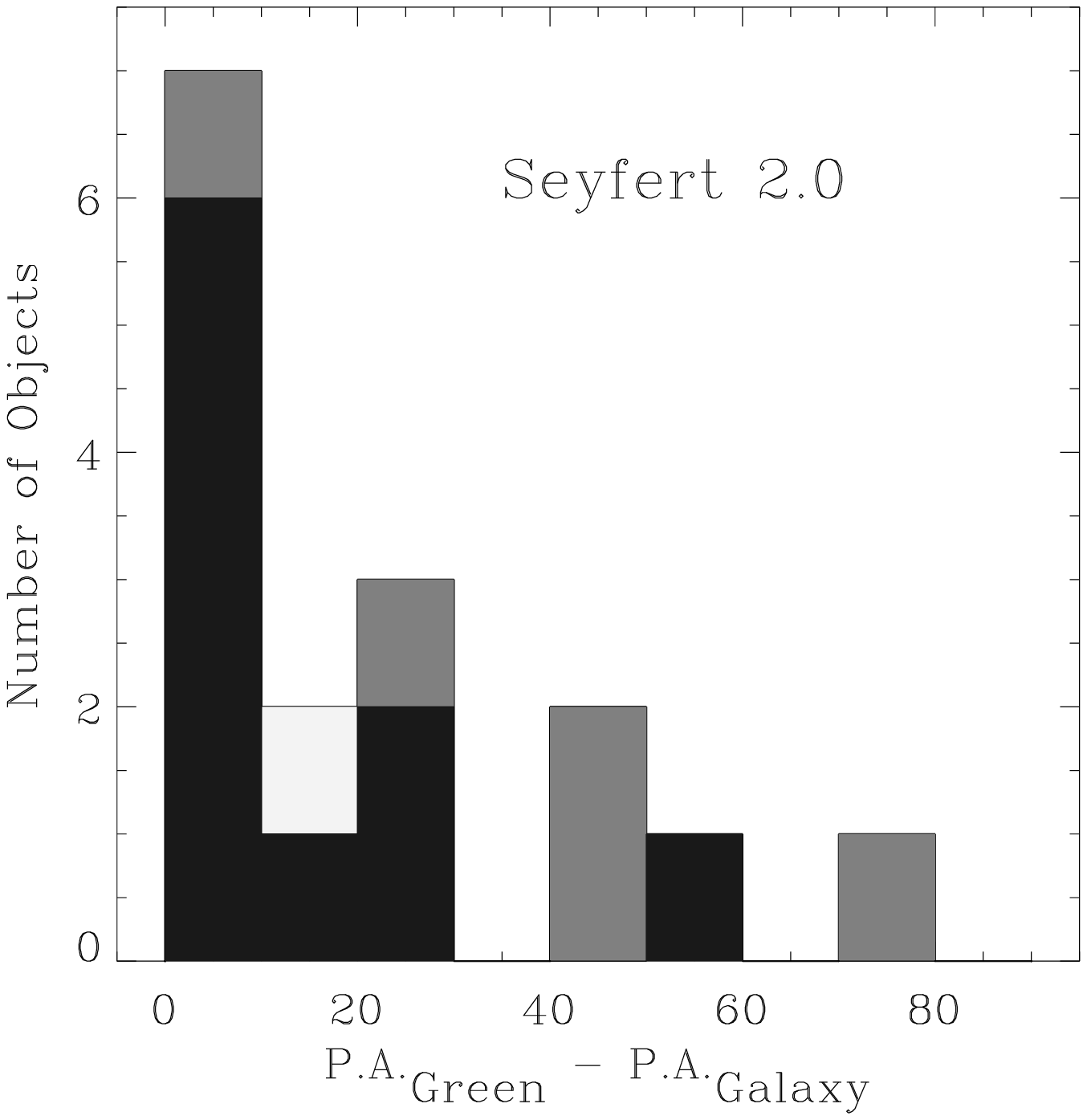}
\caption
{\small{
A comparison between two different measures (P.A.$_{Green}$ and
P.A.$_{Galaxy}$, see Sections 2.3 and 2.4)
of the P.A.'s of the major axes of the galaxies in the early-type sample.
The green continuum is measured at a surface brightness of 2~x~10$^{-18}$
ergs s$^{-1}$ cm$^{-2}$ {\AA}$^{-1}$ (arcsec)$^{-2}$, which
is the lowest contour of the images of MWZ (see Section 2.4).
Quality flags for P.A.$_{Galaxy}$ are explained in 
Section 2.3 and in Table 4 of Paper VIII. 
Quality flags for P.A.$_{Green}$ are `a', except for values in brackets in
Table~3 of Paper VIII which are assigned `b'. 
Objects with quality flag `a' for both P.A.$_{Green}$ and P.A.$_{Galaxy}$ 
are shaded black.
Objects with one flag `a' and the other flag `b' for 
P.A.$_{Green}$ and P.A.$_{Galaxy}$ are shaded dark grey. 
Objects with flag `b' for both P.A.$_{Green}$ and P.A.$_{Galaxy}$ 
are shaded light grey.
}}
\end{figure}

The data for the sample of all radio-extended Seyferts are presented in 
Tables 1, 2 and 3, with columns as follows :
(1)~galaxy name; (2)~Seyfert type.  
If broad permitted lines are present we follow, whenever possible, the
Seyfert classification scheme adopted by Whittle (1992) which
is based on the flux ratio R $=$ F$_{[OIII]}$/F$_{H\beta}$,
where F$_{[OIII]}$ is the [OIII]$\lambda$5007 flux and F$_{H\beta}$ the total
(broad plus narrow) H{$\beta$} flux.  These types are (with R values)
Sey~1.0 (R $\leq$ 0.3),
Sey~1.2 (0.3 $<$ R $\leq$ 1), Sey~1.5 (1 $<$ R $\leq$ 4),
Sey~1.8 (R $>$ 4) and Sey 1.9 (assigned if only broad
H$\alpha$ is seen). The Seyfert~1.9 subclass is unsatisfactory as it depends
on the sensitivity of the spectra. For example, NGC~5273 has been classified 
as a Seyfert~1.9 by Whittle (1992), but a deeper spectrum reveals broad wings
on H$\beta$ as well as H$\alpha$ and leads to a classification of Seyfert~1.5
(\cite{hoet97b}). Fortunately, only 6 of our galaxies are classified as 
Seyfert~1.9's.  As in Whittle (1992), some
Seyferts with highly variable H$\beta$ flux are classified as Sey 1.5.
We have used emission-line fluxes listed in Whittle (1992) and 
Winkler (1992); for objects for which we could find no measurement of R,
we use the classification given in 
``The NASA/IPAC Extragalactic Database'' 
(see e.g. Helou et al. 1991, hereafter NED). 
Whenever the Seyfert type is derived from some method other than
the flux ratio R given in Whittle (1992), the source of the
Seyfert type is noted in column 9;
(3)~radio structure, where, as in Paper V: L~= linear, D~= diffuse, 
A~= ambiguous, and S~= slightly resolved;
(4)~radio extent in kilo-parsecs (using H$_0$~= 50 km s$^{-1}$ Mpc$^{-1}$);
(5)~radio P.A., and quality flag (Section 2.2);
(6)~galaxy major axis P.A., and quality flag (Section 2.3);
(7)~inclination of the host galaxy
calculated from the photometric axis ratio of the galaxy unless indicated 
otherwise in column 9. For galaxies which appear so close to face-on 
that it is difficult to assign a value for $i$, we arbitarily list a value 
of 0{\arcdeg};
(8)~references for the [OIII] and H$\beta$ flux, the radio extent and P.A., 
and the galaxy major axis P.A. according to the key at the bottom
of Table~1; 
(9)~comments, according to the key  at the bottom of Table~1.
The appendix contains brief notes on the radio and galaxy major axes
P.A.'s for some of the galaxies listed in 
Tables 1, 2 and 3 which are not in the early-type Seyfert sample.

\section{Results}

We use the techniques of ``survival
analysis'' as coded in the ASURV software package (Lavalley, Isobe \&
Feigelson 1992) and the Kolmogorov-Smirnov (K-S) test 
to test for correlations in our data.
The numerical values of the probabilities that the distributions are drawn
from the same parent population are given in Table 4.

\vspace{-0.15in}
\subsection{Distribution of $\delta$}

Histograms of $\delta$, as measured by P.A.$_{Radio}$ $-$ P.A.$_{Galaxy}$ 
and P.A.$_{Radio}$ $-$ P.A.$_{Green}$,
for Seyferts in the early-type sample with extended radio structure and
measured galaxy major axes are plotted in Figure~3.
Histograms of $\delta$~= P.A.$_{Radio}$ $-$ P.A.$_{Galaxy}$ for the 
radio-extended Seyfert sample are shown in Figure~4.
The histograms are shaded according to the reliability of the measurement of 
$\delta$,  from black (high quality data) to
white (low quality data) - see caption to Figure~3 for details. 
There are 23 (27) galaxies in the two upper (lower) panels of Figure~3, 
and 47 galaxies in Figure~4. S97 had 46 galaxies in their study, but only 29
of them are included in Figure~4.
We did not include the remaining 17 because they do not satisfy
our criteria for a measurement of P.A.$_{Galaxy}$ (i.e.  these 17 galaxies
are round, or almost so, or distorted by interactions and were assigned
quality flag `d').  Three conclusions can be drawn from the distributions 
of $\delta$ in Figures~3 and 4: \newline
(a)~The data are severely limited by observational uncertainties.
There are very few (three Seyfert~1's, one Seyfert~1.9 and 5 Seyfert~2.0's
in Figure~4) high quality determinations of $\delta$, i.e. galaxies
with quality flag `a' for both the radio structure P.A. and
the host galaxy major axis P.A. (data shown in  black). These high quality 
values of $\delta$ still have typical errors of $\pm$5{\arcdeg}; \newline
(b)~there is no significant difference between the 
distributions of $\delta$ for Seyfert~1's and Seyfert~2.0's (Table~4).
The statistical tests were
applied to the complete distribution (all quality flags) and 
also to all subsets of the complete distribution formed by selecting
only those data better than a certain quality level; \newline
(c)~the $\delta$ distribution of the Seyfert~2.0's is different from a
uniform distribution at the $\sim$90\% confidence level, while the  
$\delta$ distribution of the Seyfert~1's is not significantly 
different from a uniform distribution (Table~4). As found by S97, there seems
to be a lack of Seyfert~2.0's with $\delta$~$>$~70{\arcdeg}. A posteriori
Poisson statistics give a 3\% probability of finding only the one 
Seyfert~2.0 at $\delta$~$>$~70{\arcdeg} (Fig. 4) instead of the expected five.
\newline
Including the Seyfert~1.8 and 1.9 galaxies in either the Seyfert~1 or the
Seyfert~2.0 distribution does not change result (b). If we add the 
Seyfert~1.8's and 1.9's to the Seyfert~2.0's, the resulting distribution
of $\delta$ is different from uniform at a confidence
level of only $\sim$80\%--90\%.

\newpage

\begin{figure}[!ht]
\figurenum{3}
\plottwo{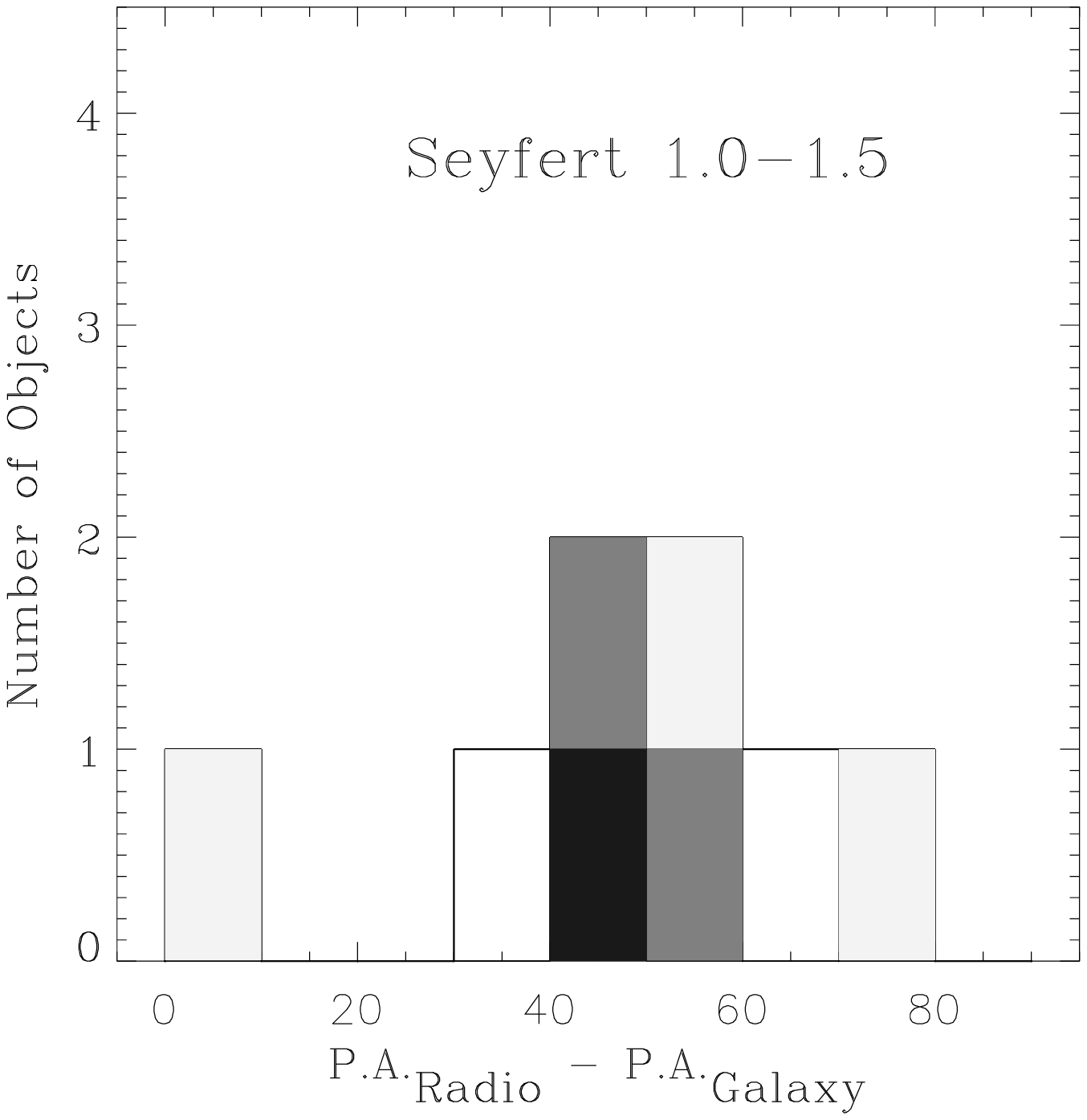}{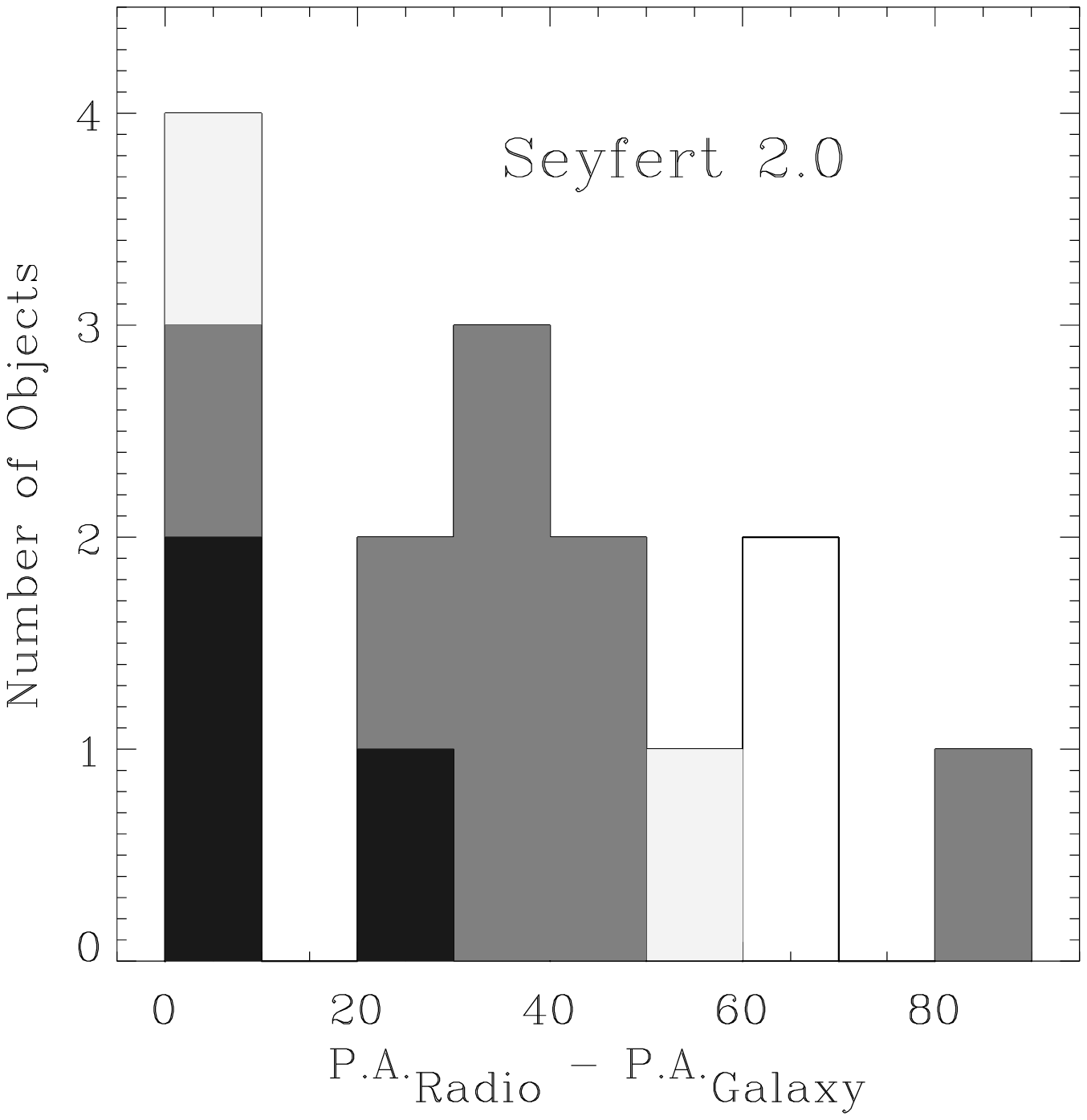}

\plottwo{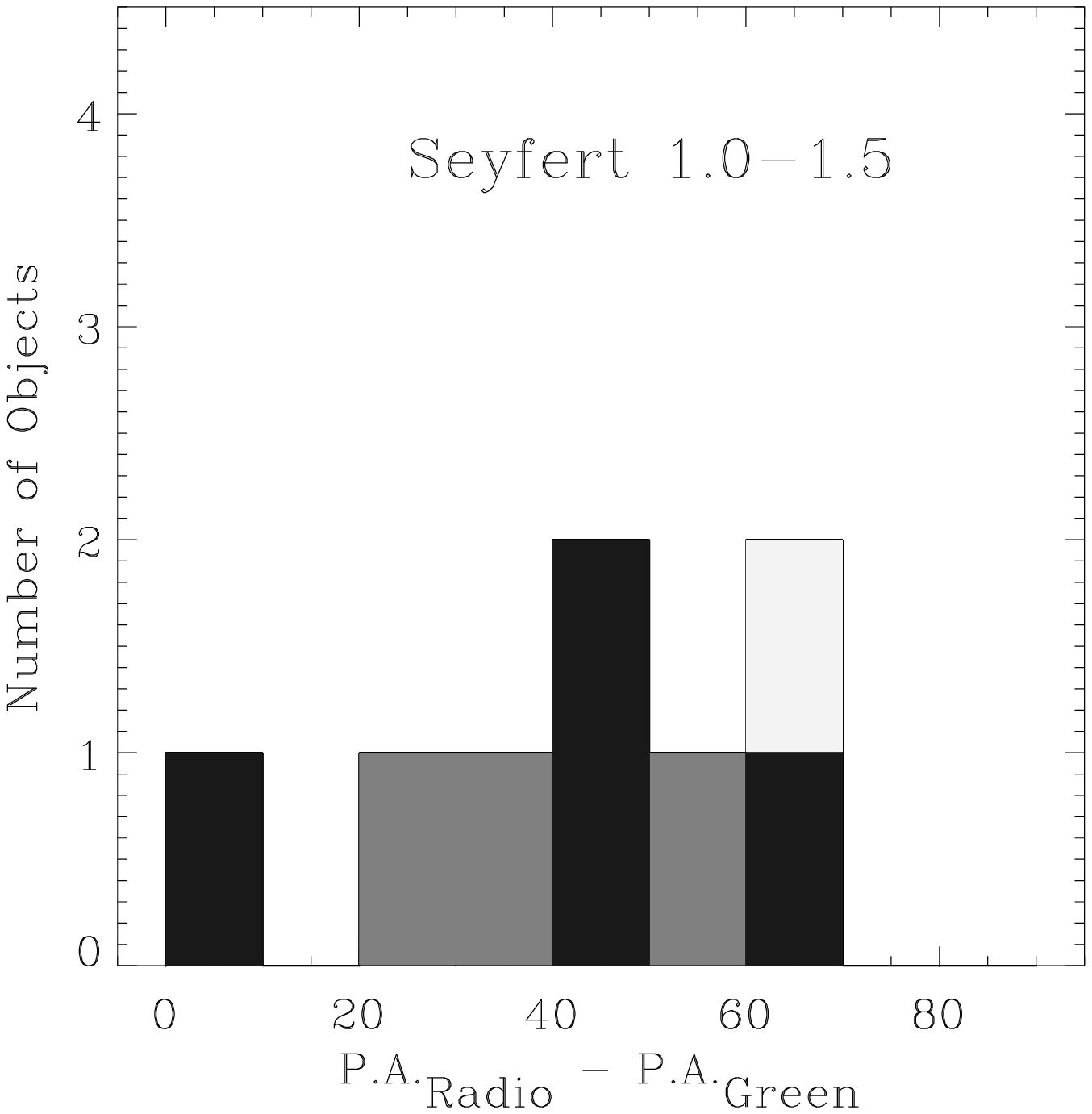}{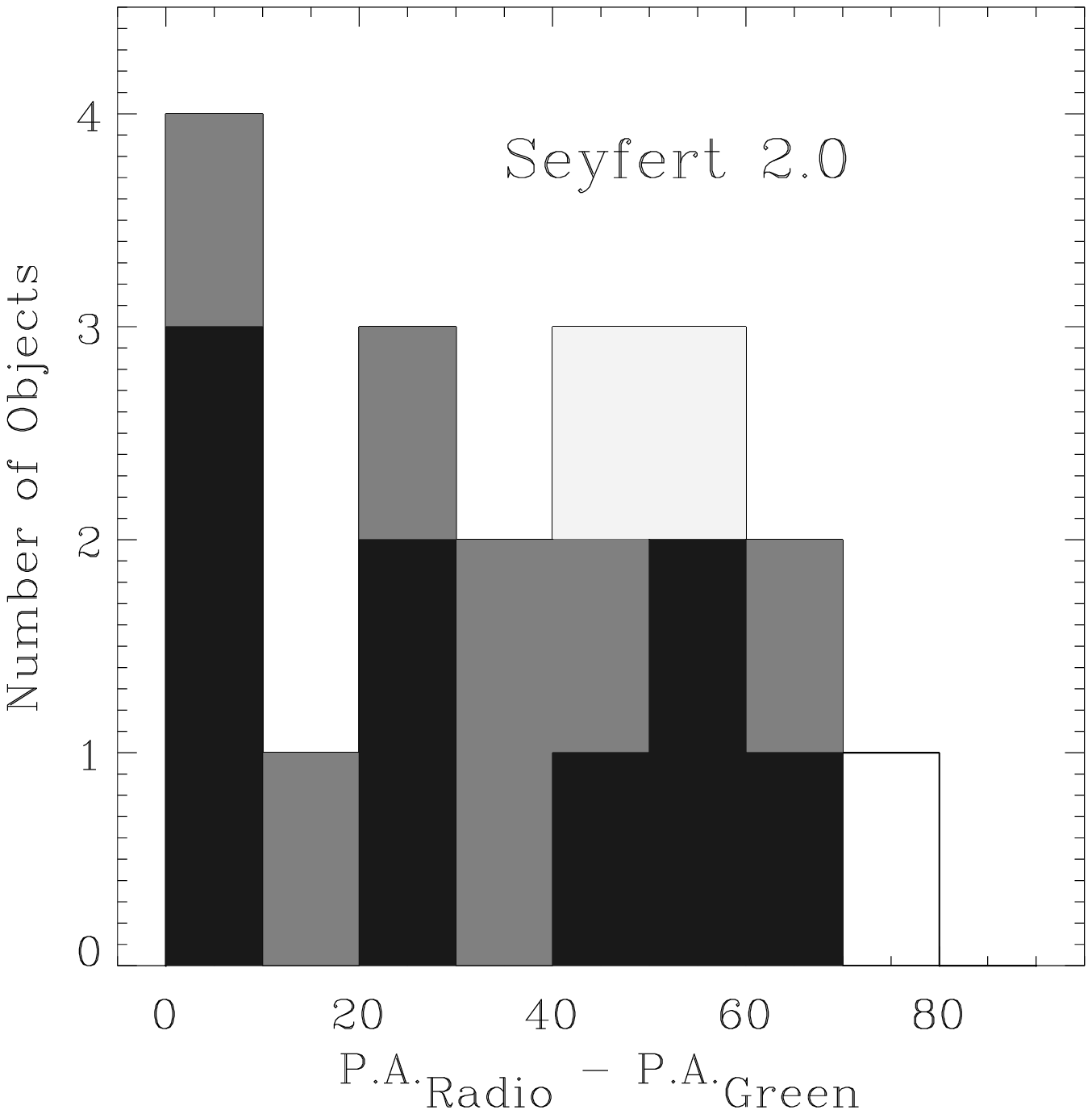}
\caption
{\footnotesize{ 
Histograms of $\delta$ as measured by
P.A.$_{Radio}-$P.A.$_{Galaxy}$ and P.A.$_{Radio}-$P.A.$_{Green}$ 
for the early-type Seyfert sample. 
Quality flags for P.A.$_{Radio}$ and P.A.$_{Galaxy}$ are explained in 
Sections 2.2 and 2.3, and in Table 4 of Paper VIII. 
Quality flags for P.A.$_{Green}$ are `a', except for values in brackets in
Table~3 of Paper VIII which are assigned `b'. 
Objects with quality flag `a' for both P.A.$_{Radio}$ and P.A.$_{Galaxy}$
(or P.A.$_{Green}$) are shaded black. Objects with one quality flag `a' and
the other flag `b' for P.A.$_{Radio}$ and P.A.$_{Galaxy}$ (or P.A.$_{Green}$)
are shaded in dark grey. Objects with  flag `b' for both P.A.$_{Radio}$ and 
P.A.$_{Galaxy}$ (or P.A.$_{Green}$) are shaded light grey. All others are 
white.
Top~left:~distribution of P.A.$_{Radio}-$P.A.$_{Galaxy}$ for Seyfert~1's
(Seyfert~1.0's through Seyfert~1.5's).
Top~right:~distribution of P.A.$_{Radio}-$P.A.$_{Galaxy}$ for Seyfert~2.0's.
Bottom~left:~distribution of P.A.$_{Radio}-$P.A.$_{Green}$ for Seyfert~1's
(Seyfert~1.0's through Seyfert~1.5's).
Bottom~right:~distribution of P.A.$_{Radio}-$P.A.$_{Green}$ for Seyfert~2.0's.
}}
\end{figure}

\newpage

\begin{figure}[!ht]
\figurenum{4}
\plottwo{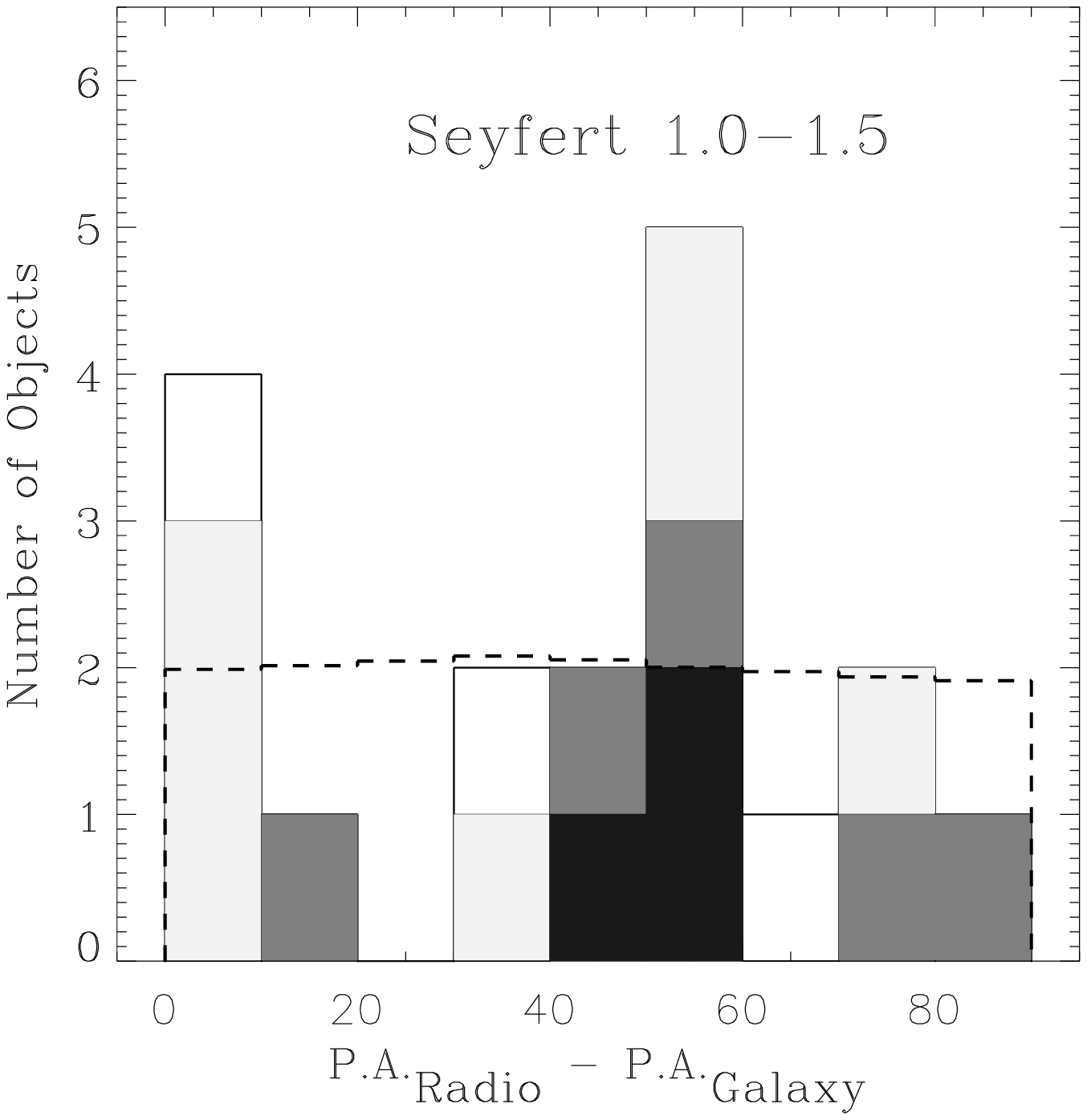}{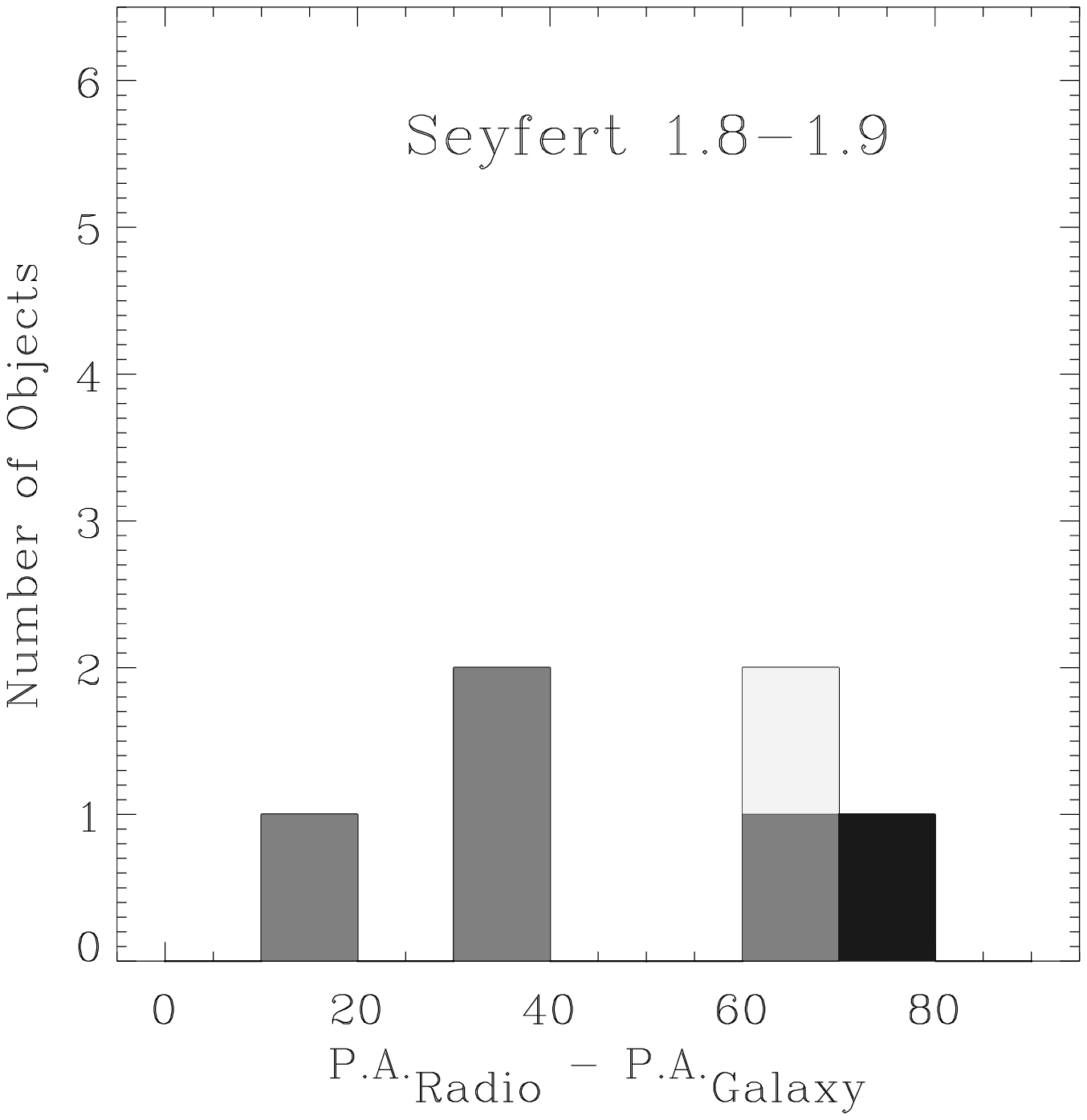}

\plottwo{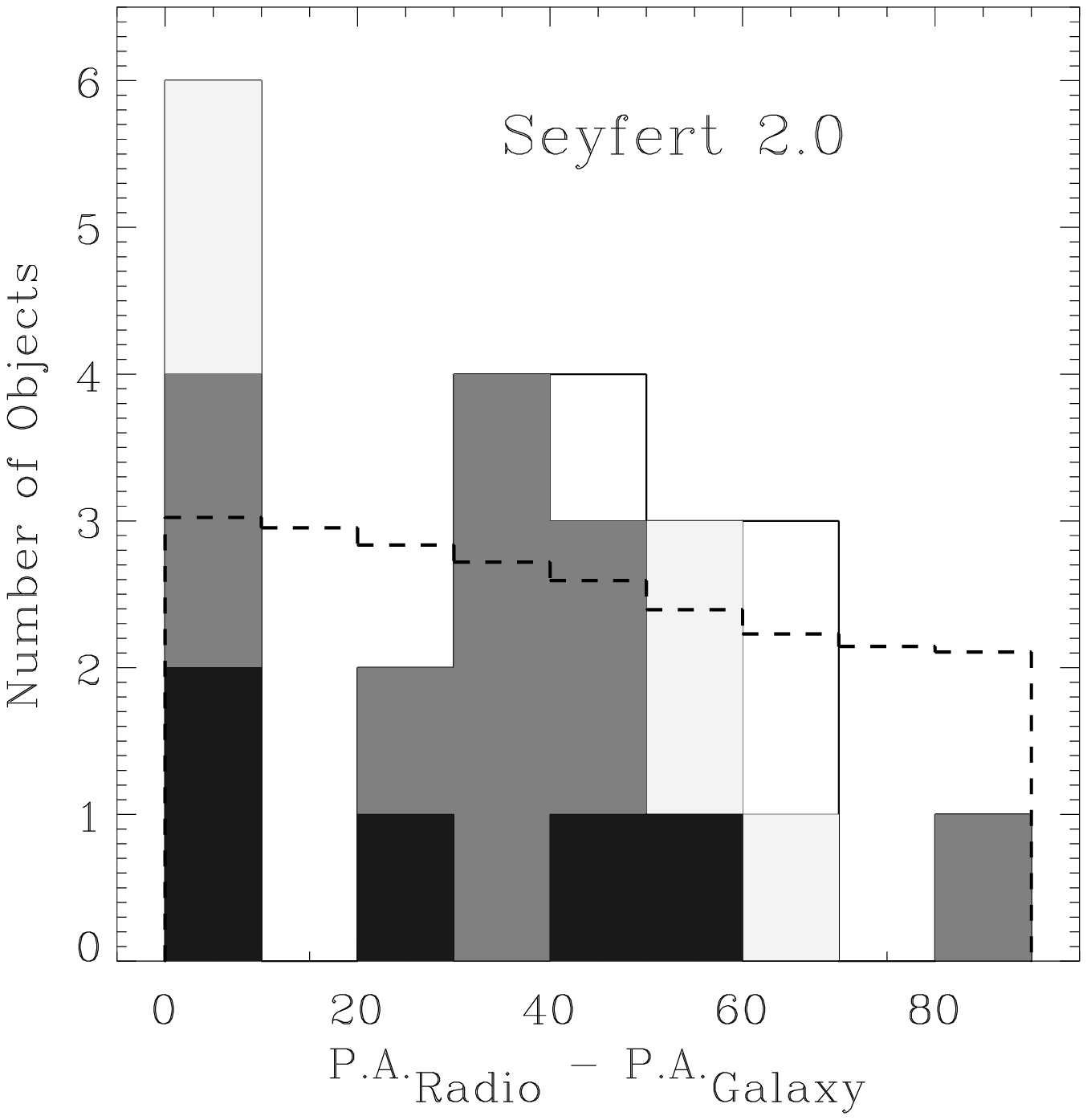}{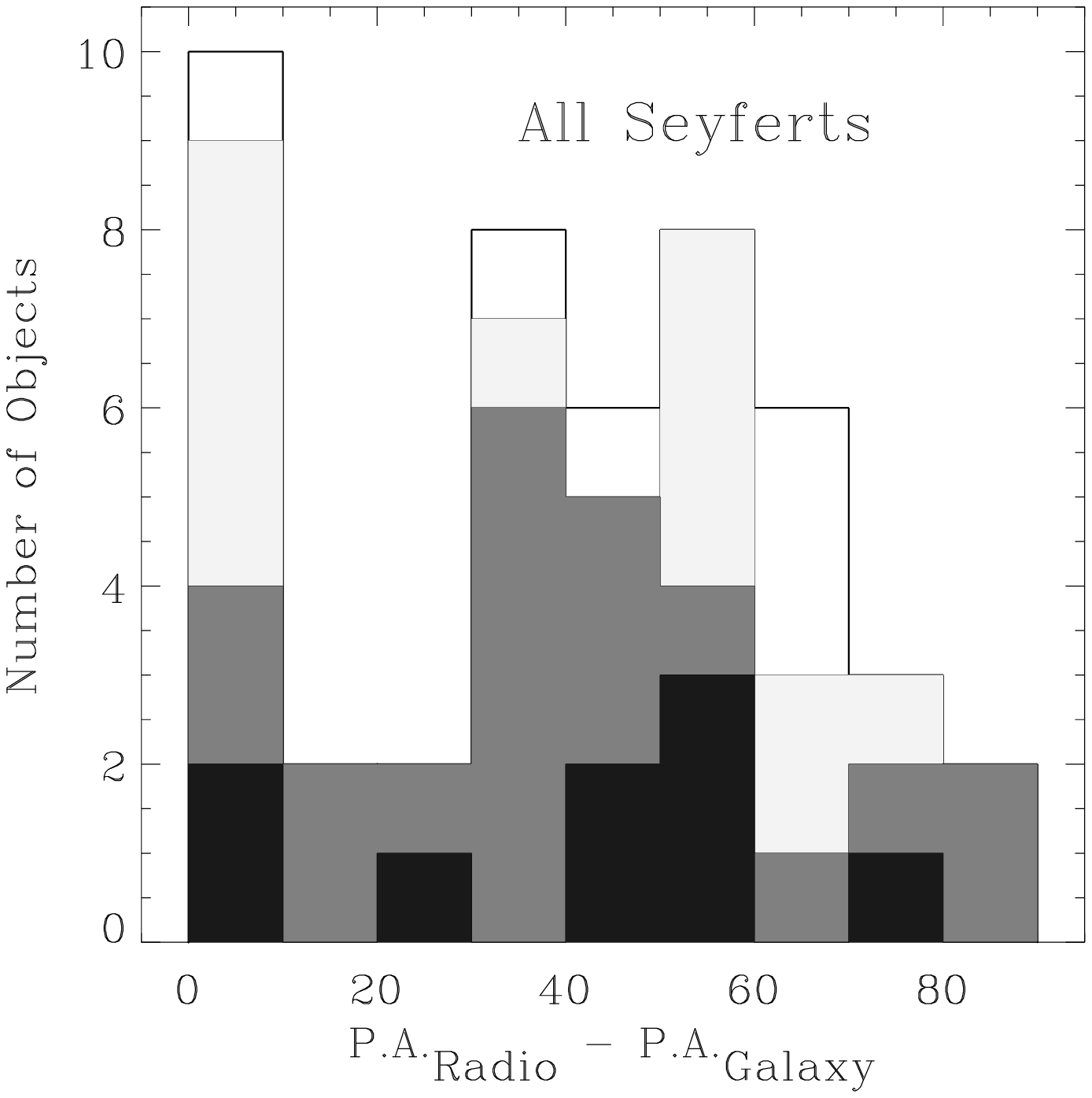}
\caption
{\footnotesize{
Histograms of  $\delta$~=~P.A.$_{Radio}-P.A._{Galaxy}$
for all Seyferts in the sample of radio-extended Seyferts.
The greyscale coding is as given in the caption to Figure~3.
Top~left:~Seyfert~1's (Seyfert~1.0's through Seyfert~1.5's). 
The thick dashed line shows the expected distribution of $\delta$ for a
$\beta$ distribution which favors higher values of $\beta$ (see Section~4.1).
Top~right:~Seyfert~1.8's and Seyfert~1.9's.
Bottom~left:~Seyfert~2.0's.
The thick dashed line shows the expected distribution of $\delta$ for a
$\beta$ distribution which favors higher values of $\beta$ (see Section~4.1).
Bottom~right:~all Seyfert types.
}}
\end{figure}

\begin{figure}[!ht]
\figurenum{5}
\plottwo{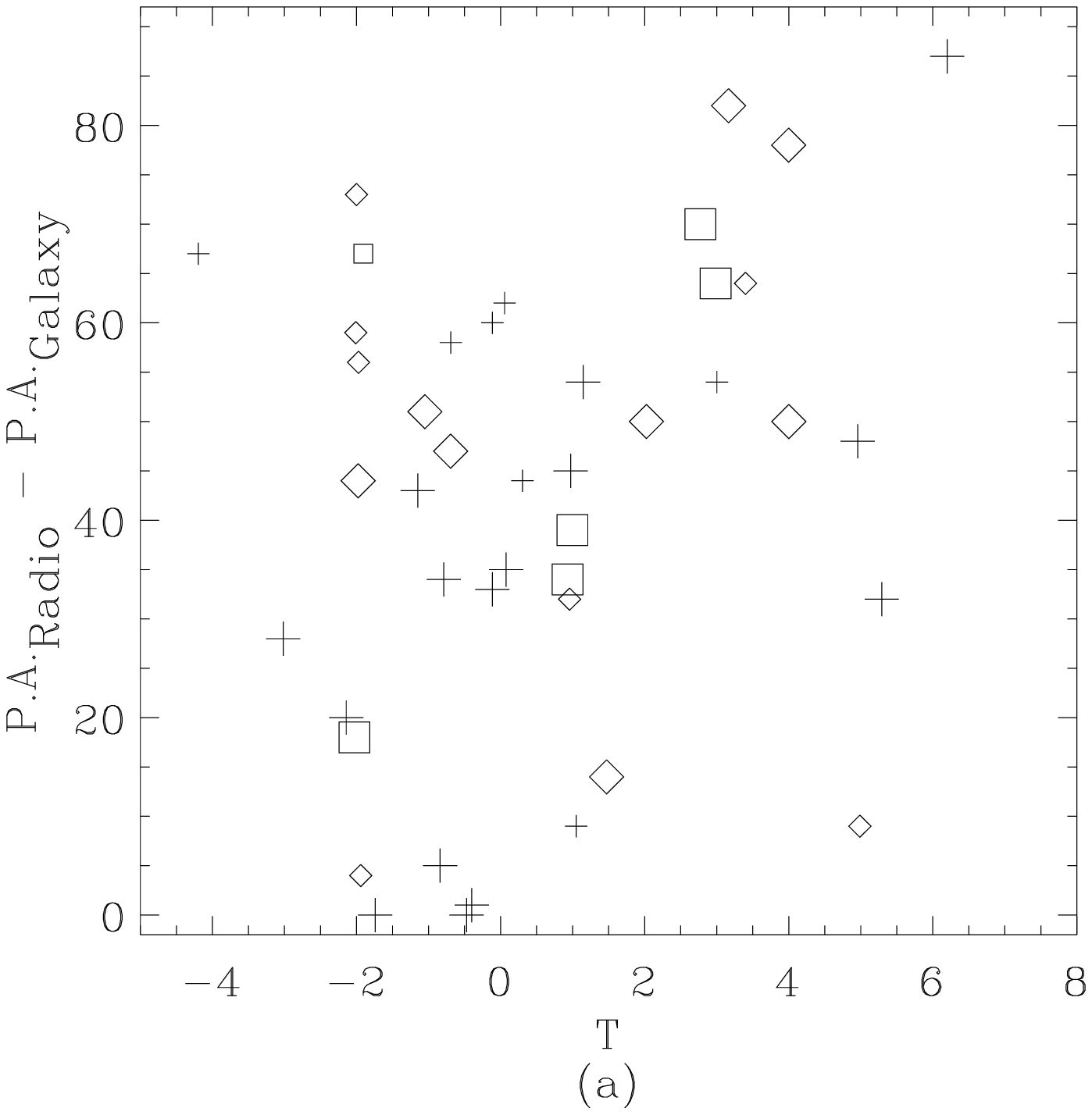}{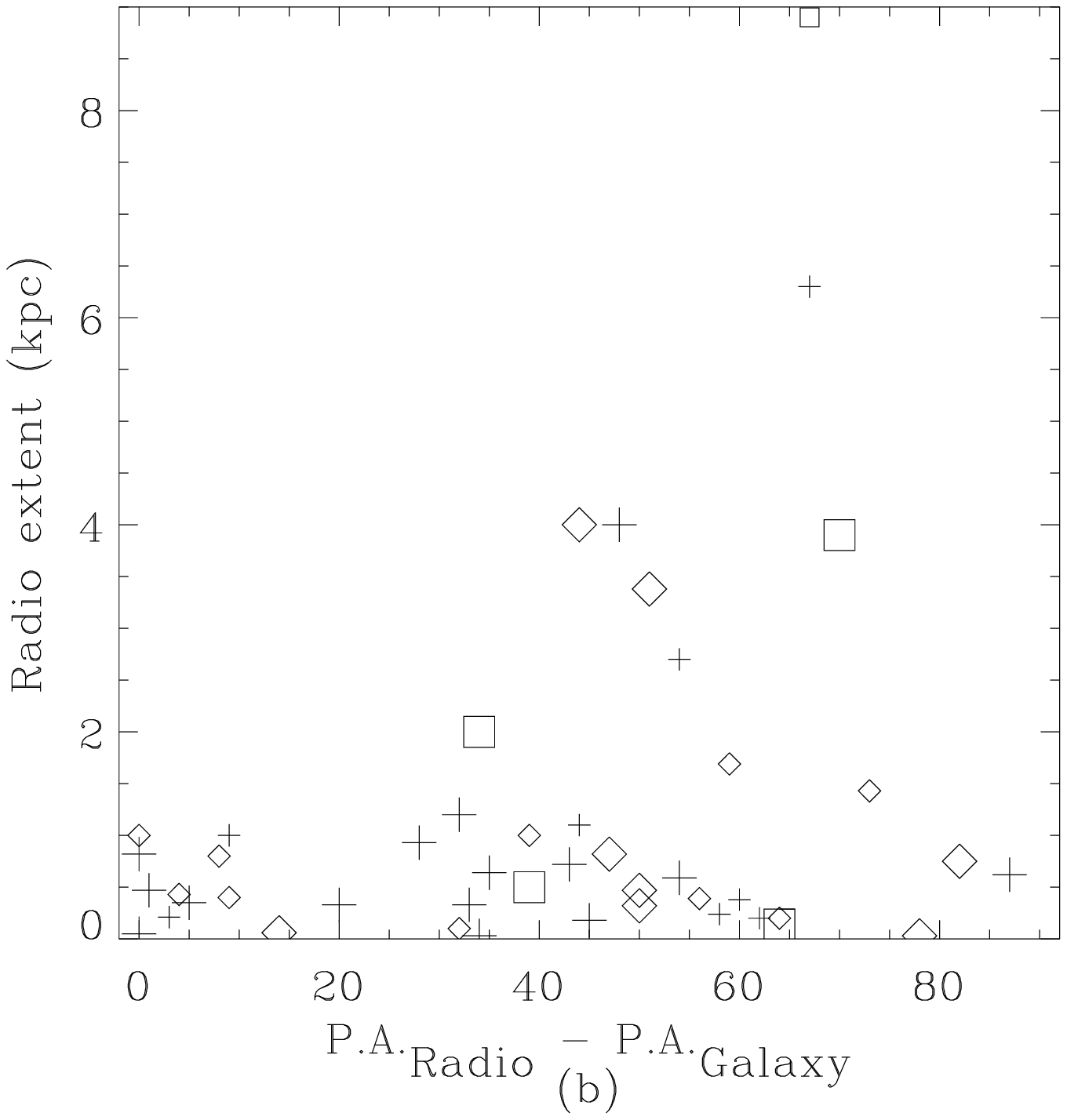}
\caption
{\small{
The relationship between $\delta$~=~P.A.$_{Radio}-P.A._{Galaxy}$, 
host galaxy morphological type, and radio extent.
Seyfert~1's (Seyfert~1.0's through Seyfert~1.5's)
are plotted as diamonds; Seyfert~2.0's are plotted as
crosses; Seyfert~1.8's and Seyfert~1.9's are plotted as squares.
Large symbols denote higher quality data.
Due to its large extent, NGC~5252 (14.8~kpc)  does not appear in Figure~5b.
(a)~Relationship between $\delta$ and galaxy morphological type, as
defined in RC3;
(b)~relationship between radio extent and $\delta$.
}}
\end{figure}
Seyferts in host galaxies with morphological type earlier than Sab 
(T~=~2) show a distribution of $\delta$ that is more or less 
uniformly distributed, with the exception of the possible deficiency for 
$\delta~>$~70{\arcdeg} noted above (Fig.~5a and Table~4). 
However, Seyferts in galaxies with host type Sab and later may favor 
larger values of $\delta$ (at a $\sim$90\% confidence level; Table~4). 
The only late-type Seyfert with a low value of $\delta$,
Mrk~231, is a well known system of two galaxies in an 
advanced merger stage\footnote{
The outer isophotes of the DSS image
of this galaxy do not show obvious signs of the merger, so it was
retained in the sample for consistency.}
(\cite{sanet87}). 
The remaining late-type Seyferts
do not show obvious signs of on-going mergers.
Without Mrk~231, the distribution of $\delta$ for Seyferts of host type
Sab and later is different from uniform at the $\sim$96\% confidence level.
Since a distribution of $\beta$ which favors low values gives 
a distribution of $\delta$ that favors large values,
this result suggests that the nuclear accretion disk in non-interacting
late-type spirals
shares the angular momentum direction of the stellar disk. However, a 
larger sample is needed for a conclusive result. In early-type
galaxies, the accretion disk is more randomly oriented with respect to the
stellar disk and may have resulted from accretion following a galaxy merger.
There is no significant correlation between the radio extent and $\delta$
(Fig.~5b), though it appears that the maximum extent at a specific value of
$\delta$, increases with increasing $\delta$.

\subsection{The 3-D geometry from $\delta$ and $i$} 
The distribution of $\delta$ (Figs.~3 and 4) shows the relationship between
the radio axis and host galaxy major axis, as projected on the sky.
C98 have demonstrated
that knowledge of both $\delta$ and $i$, the host galaxy 
inclination, can be used in a 3-dimensional statistical analysis to 
constrain the values of $\beta$ and $\phi$ (see Fig.~1 for a definition of
the angles).
Briefly, they show that for a given pair of values of $i$ and 
$\delta$, the jet vector is constrained to lie on a great circle of a
sphere centered on the galaxy. This great circle passes through the line
of sight and lies at some calculable angle with the plane of the galaxy. The 
value of $\beta$ depends on the exact location of the jet vector on the great
circle, which is not known.
One can, however, calculate the minimum possible value of $\beta$ which
corresponds to the case where the great circle passes closest to the 
galaxy disk axis. C98 show that  \newline
$\beta_{min}~= \arccos(\sin^2{\delta}~\sin^2i +
	\cos^2i)^{\onehalf}$.  

\begin{figure}[!t]
\figurenum{6}
\plottwo{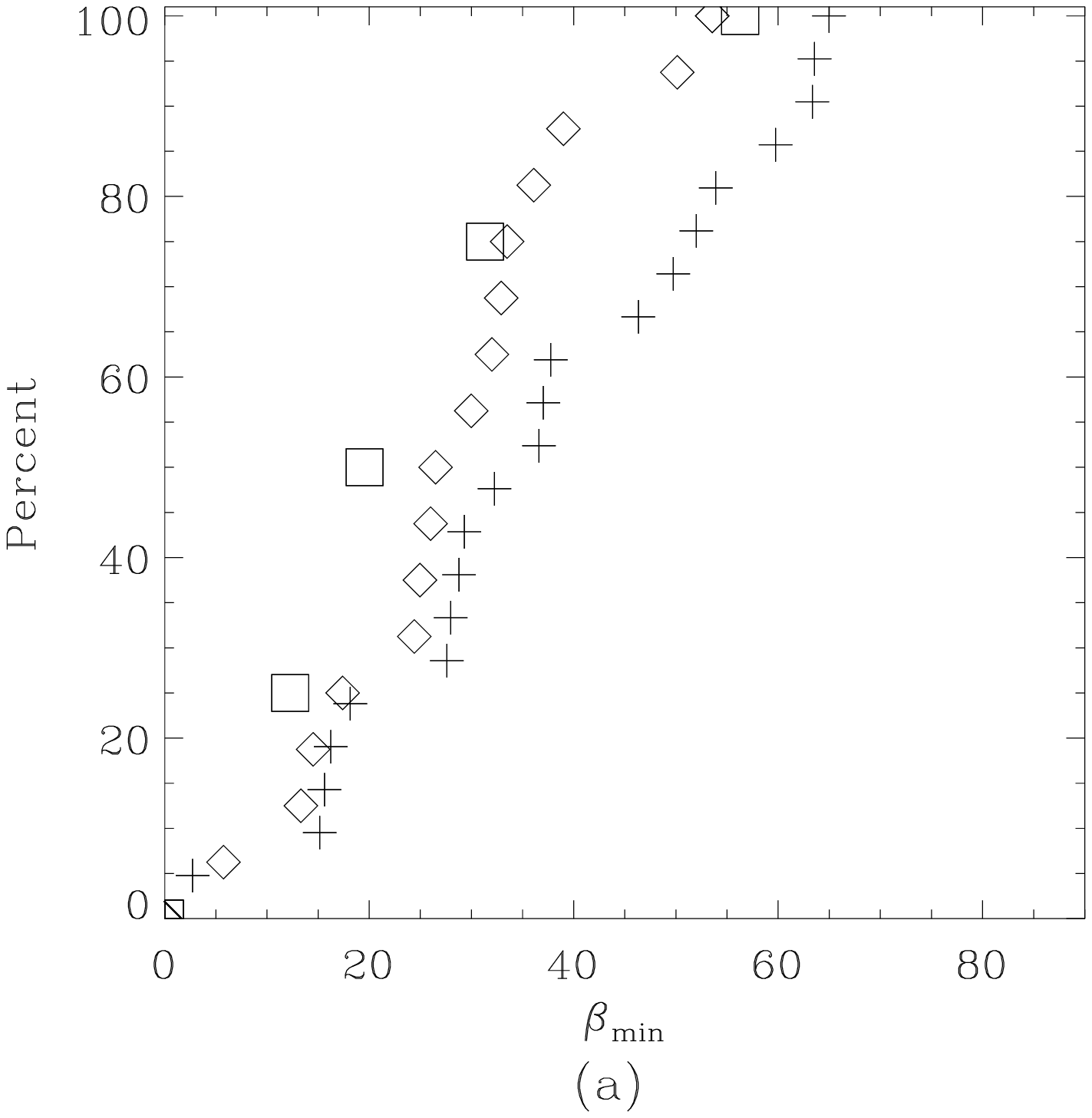}{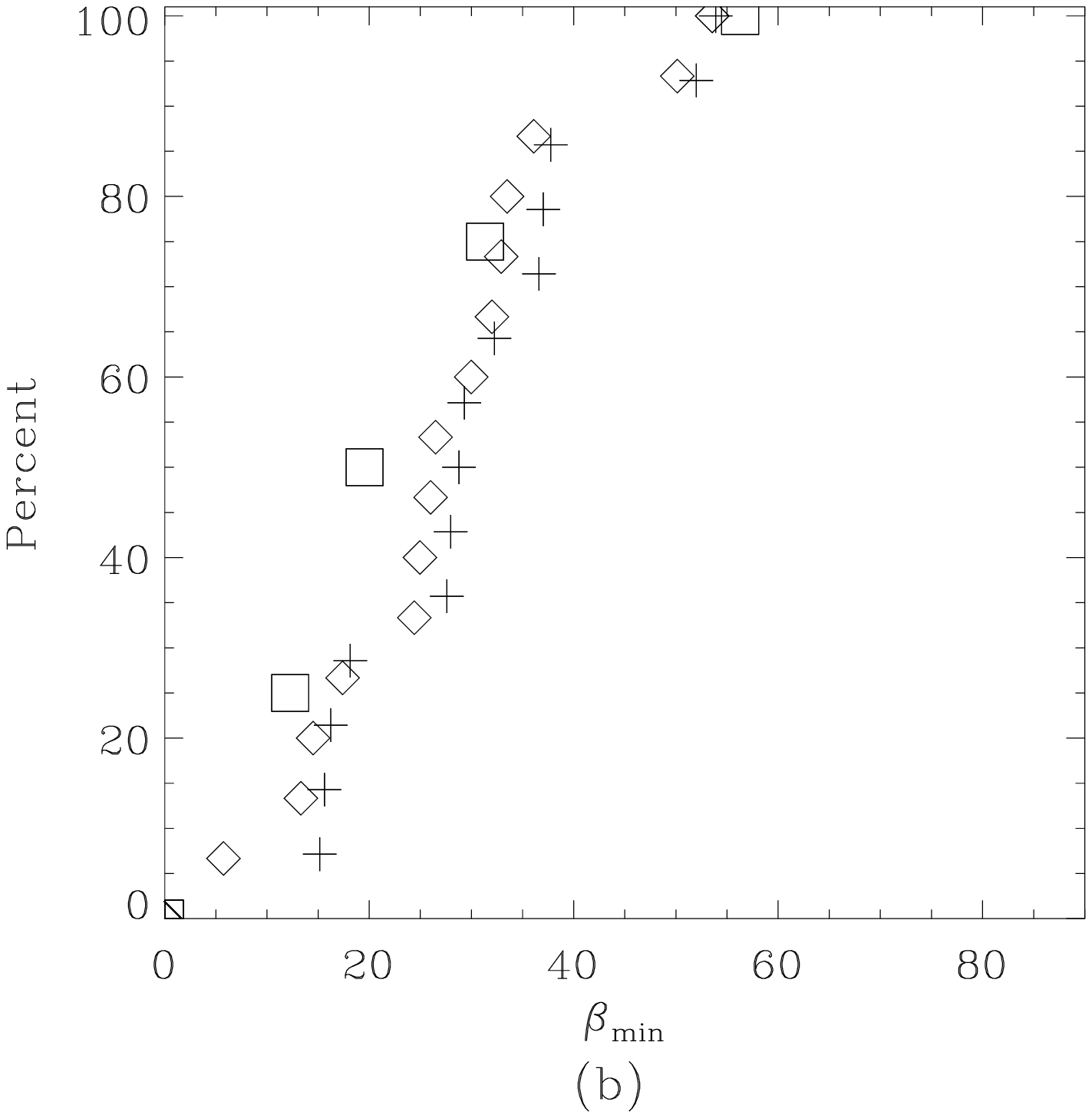}
\caption
{\small{
The cumulative distribution of $\beta_{min}$ for the radio-extended Seyferts.
Seyfert~1's (Seyfert~1.0's through Seyfert~1.5's) are plotted as diamonds;
Seyfert~2.0's are plotted as crosses; Seyfert~1.8's and 1.9's are plotted 
as squares.
The three Seyfert distributions have been normalized and are plotted as a
percentage. 
(a) All radio-extended Seyferts which have well-defined values
of $\delta$ and $i$;
(b) same as (a), except Seyfert's with 
$i~\geq$~60{\arcdeg} are omitted.
}}
\end{figure}

We have calculated the values of $\beta_{min}$ for all Seyferts in the
radio-extended Seyfert sample for which values of both $\delta$ and
$i$ are available and the results (Fig.~6a) are similar to those of C98,
namely Seyfert~1's appear to favor smaller values of $\beta_{min}$ at a 
confidence level of $\simeq$~90\% (Table~4).
The difference between the two $\beta_{min}$ distributions appears larger
($\simeq$~95\% confidence level, Table~4)
when we consider only higher quality data (quality flag `b' or better),
though the Kolmogorov-Smirnov (K-S) test only supports the difference at
the 82\% confidence level.
This difference, however, is probably biased by the difference in the 
inclination distributions of the Seyfert~1's and 2.0's.
Though the distributions of $i$ for Seyfert 1's and 2.0's in our sample are
not significantly different in a statistical sense (Table~4), Seyfert~2.0's
are seen in higher inclination host galaxies than Seyfert~1's (Fig.~7).
For example, eight (27\%) Seyfert~2.0 host galaxies have $i~>~60{\arcdeg}$ as
compared to only one (5\%) Seyfert~1 host galaxy. 
Since $\beta_{min}$ can also be expressed as  \newline
$\beta_{min}~= \arccos(1 - \sin^2i~\cos^2\delta)^{\onehalf}$ \newline
smaller values of $i$ will result in smaller values of $\beta_{min}$ for
a given $\delta$.
Indeed, omitting all galaxies with $i~>$~60{\arcdeg} from the sample
results in similar $\beta_{min}$ distributions for Seyfert~1's and 2.0's 
(Fig.~6b and Table~4). We conclude there is no significant difference
between the distributions of $\beta_{min}$ for Seyfert 1's and 2.0's.

\begin{figure}[!ht]
\figurenum{7}
\plotfiddle{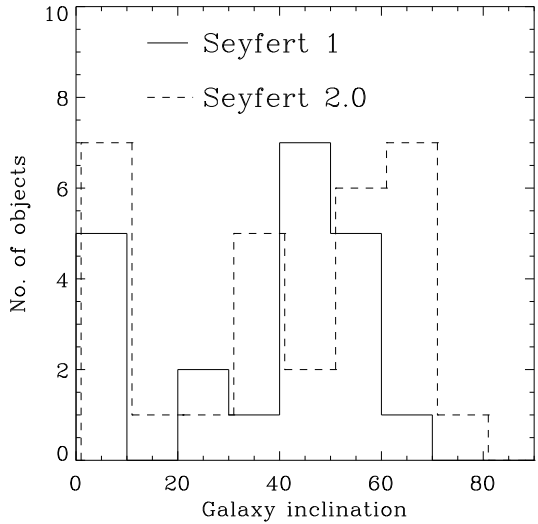}{3in}{0}{50}{50}{-180}{-80}
\vspace{-0.5in}
\caption
{\small{
The distribution of host galaxy inclination for the radio-extended 
Seyferts. The Seyfert~2.0 distribution has been slightly
offset along the X axis for clarity. 
A larger fraction of the Seyfert~2.0's than the Seyfert~1's
are found in high inclination ($i~>$~60{\arcdeg}) galaxies.
Galaxies with $i~\leq$~20{\arcdeg} are, in general, too
face-on to determine a value of P.A.$_{Galaxy}$ (and hence $\delta$)
or $i$; these galaxies have been assigned value $i$~=~0{\arcdeg}.
}}
\end{figure}

We have also estimated the probability distribution of $\beta$
for all Seyfert types using the method developed by C98.
In this method, one starts with an assumed probability distribution for
$\beta$, P$_{in}(\beta)$, such as the hypothesis that the jets are randomly
oriented, P$_{in}(\beta)$~=~$\sin\beta$. Then, for each data pair 
($\delta$,$i$), one 
calculates the range of possible values for $\beta$ and $\phi$.
The data point is then distributed, using appropriate weighting, over all 
these possible values of $\beta$ and $\phi$. After distributing all data
points, one can integrate over azimuth ($\theta$; see Fig. 1) and recover an 
output probability distribution for $\beta$, P$_{out}(\beta$).
If P$_{in}(\beta)$ and P$_{out}(\beta)$ are similar,
then the input distribution is a good approximation to the actual
distribution of $\beta$.  In agreement with C98,  we find that 
the ($\delta$,$i$) data for the radio-extended Seyfert sample are consistent
with the idea that the radio jets are randomly distributed with respect to 
the host galaxy plane when all Seyfert types are taken together (Fig.~8a).
This results still holds when we use only higher quality data (quality flag 
`b' and above). There is no significant difference between the distributions
of $\beta$ for Seyfert~1's and 2.0's when these are considered separately
(Figs. 8b and 8c).

\begin{figure}[!ht]
\figurenum{8}
\plotone{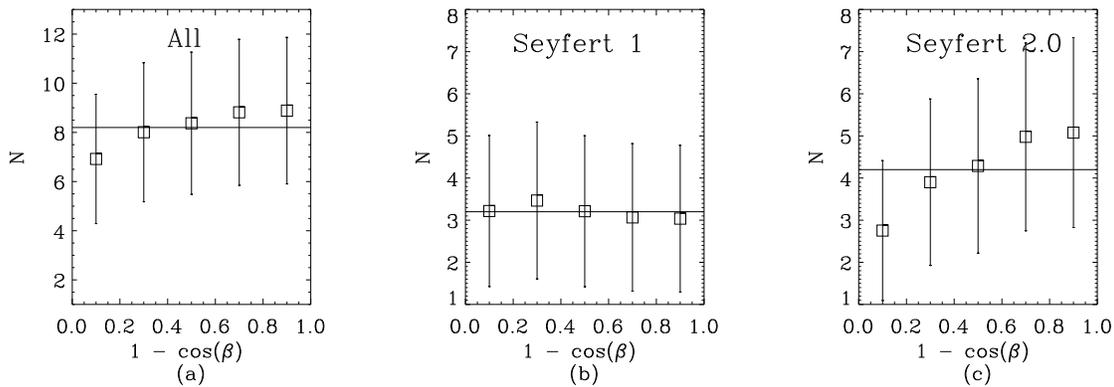}
\vspace{-6.3in}
\caption
{\small{
The estimated P($\beta$) distribution obtained for Seyferts in
the radio-extended sample, using the method described by C98.
The bins are chosen to have equal sizes in
1~$-~\cos(\beta)$ so that a random distribution of jets in space would
give equal numbers in each bin. The error bars have values 
$\sqrt{n}$, where $n$ is the number of objects in each bin. 
The solid line shows the expected behavior for a random distribution of jets.
Estimated P($\beta$) distribution for 
(a)~Seyferts of all types, 
(b)~Seyfert~1's,
and (c)~Seyfert~2.0's.
}}
\end{figure}

Further, following C98, we can assume that the unified model is valid, 
namely that Seyfert~1's and Seyfert~2's are distinguished by whether $\phi$ is
smaller or greater than, respectively, $\phi_c$. We then obtain the $\beta$
distributions shown by the data points in Figure~9 for various assumed values
of $\phi_c$. The results for $\phi_c$~=~40{\arcdeg} are similar to those 
calculated by C98 (their Fig.~5), with an apparent trend for Seyfert~2's
to favor high values of $\beta$ and Seyfert~1's low values of $\beta$. 
As we now show, this apparent difference between the Seyfert~1 
and Seyfert~2.0 $\beta$ distributions is expected even if the radio jets
are randomly distributed with respect to the host galaxy disk, i.e. even if 
the 
actual $\beta$ distribution is uniform in $\cos\beta$ for both types of
Seyferts.
For galaxy inclination $i$, a jet with $\beta~=~i$
and $\theta~=~-$90{\arcdeg} will be directed along the l.o.s. and the 
galaxy will be seen as a Seyfert~1. In fact, all galaxies with  
jets within $\phi_c$ of the l.o.s. will be observed as Seyfert~1's, so for
this type of Seyfert, values of $\beta$ close to $i$ will be preferentially 
selected. 
The distributions of $i$ for the radio-extended Seyferts show a deficiency of 
high values of $i$ (Fig.~7), which is expected in a sensitivity-limited
Seyfert sample due to obscuration in the galaxy disk (e.g. Ho, Filippenko
\& Sargent 1997a). This deficiency is especially strong for Seyfert 1's.
For the present analysis, there is also a lower $i$ cutoff of 
$\sim$20{\arcdeg}, since nearly face-on galaxies (major to minor axes ratio
$\lesssim$~1.1) are
omitted from the sample because they do not have a well defined
galaxy major axis (see Section 2.3). 
This deficiency of high (low) values of $i$ results in a deficiency of high 
(low) values of $\beta$ for Seyfert~1's. The rightmost bin in each 
panel of Figure~9 
consists of 71{\arcdeg}~$<~\beta~<$~90{\arcdeg}. Seyfert~1 galaxies
in this bin have inclinations $i~>~71{\arcdeg}-\phi_c$ and so should be 
relatively few in number. The actual number is very sensitive to the  
upper inclination cutoff in the observed sample and the value of $\phi_c$.
The depletion of Seyfert~1's in the leftmost bin is not as
significant because the surface area of the hemisphere between $\beta$~=
0{\arcdeg} and 20{\arcdeg} is $<$~0.2 times the surface area  
between $\beta$~= 71{\arcdeg} and 90{\arcdeg}. 
The expected distributions of $\beta$ for Seyfert~1's and 2.0's, using
the actual observed distributions of $i$ for the sample (Fig.~7),
and the assumption that $\beta$ is randomly distributed,
are shown by the solid lines in Figure~9. These expected $\beta$ distributions
for values of $\phi_c~\sim 30{\arcdeg}-50{\arcdeg}$ do not change
significantly if, instead, we use a distribution of $i$
that is uniform in $\cos~i$ over the range 20{\arcdeg}~$<~i~<$~65{\arcdeg}
and zero elsewhere.
For values of $\phi_c~\lesssim~$30{\arcdeg}, the expected $\beta$ distribution
for Seyfert~2.0's is barely different from uniform in $\cos\beta$, as 
Seyfert~2.0's are seen over most (0.87 if $\phi_c$=30{\arcdeg}) 
of the hemisphere.
A simple $\chi^2$ distribution (with two degrees of freedom) shows no
significant difference between the observed and expected distributions
of $\beta$ for both Seyfert~1's and Seyfert~2.0's. 
We also find that the data cannot be used for constraining $\phi_c$ 
as the observed and expected distributions of $\beta$ are in agreement for a 
wide range of values of $\phi_c$ (Figure~9).

The inclination bias that we have discussed here does not affect the expected
$\delta$
distributions of Seyfert~1's and 2.0's. If the jets are randomly oriented,
then, regardless of the galaxy inclination distribution, the expected
distribution of $\delta$ is uniform for both Seyfert types.
The weak trend for the distribution of $\beta$ for Seyfert~2.0's to 
increase with $\beta$ is a consequence of the weak trend for the
distribution of $\delta$ to decline toward higher $\delta$
for these galaxies (Section 3.1 and 4.1).

\begin{figure}[!ht]
\figurenum{9}
\plotfiddle{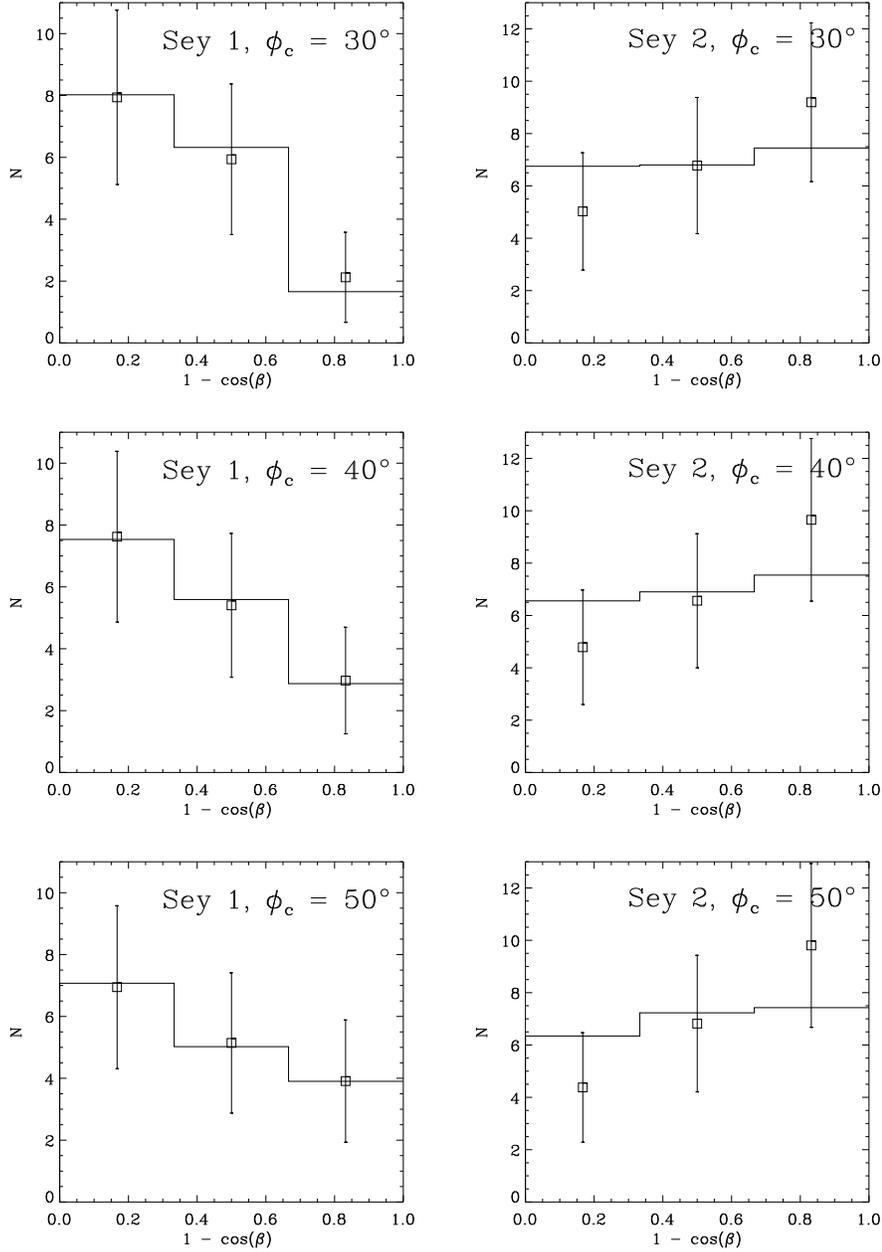}{7in}{0}{75}{75}{-250}{0}
\vspace{-0.8in}
\caption
{\small{
The estimated P($\beta$) distribution obtained for Seyferts in
the radio-extended sample, for various values of $\phi_c$, under the 
assumption that it is only the angle $\phi$ which distinguishes 
Seyfert~1's from 2.0's.
The bins are chosen to have equal sizes in
1~$-~\cos(\beta)$ so that a random distribution of jets in space with a 
random orientation of host galaxy disks would
give equal numbers in each bin. The error bars have values 
$\sqrt{n}$, where $n$ is the number of objects in each bin. 
The solid lines are {\it{not}} a fit to the data, but instead show the expected
behavior for a random distribution of jets
and the observed distributions of host galaxy inclinations (see Section 3.2).
}}
\end{figure}

There is no known relativistic beaming in Seyferts on the scale of our
radio extents (hundreds of pc~--~ several kpc), so the brighter radio jet
may point towards ($\mid\phi\mid~<~90{\arcdeg}$; Fig. 1) or away 
($\mid\phi\mid~>~90{\arcdeg}$) from us.
Further, there is no significant absorption in the disk at
short centimeter wavelengths,
so the brighter jet may be on the near side ($Z~>$~0; Fig. 1) or the
far side
($Z~<$~0) of the galaxy disk. For the analysis in this paper, symmetry ensures
that it is sufficient to consider the region $Z~>$~0 only (C98).
The relative geometry in NGC~1068, and perhaps NGC~3516, can be more closely
constrained. In NGC~1068, the HI kinematics (\cite{brimun96})
and the assumption of trailing spiral arms indicate that the nearer half
($Y~<~0$; Fig. 1) of the galaxy disk is to the S. 
The NE radio-lobe is located in front of the
HI disk and the SW radio-lobe behind (\cite{galet94}), so we know
that the NE radio lobe projects 
against the farther side of the galaxy and has $\phi~>~0{\arcdeg}$.
In NGC~3516, Ferruit, Wilson \& Mulchaey (1998) present evidence that the
nearer half of the galaxy disk is to the NW, and that the ionized-gas
associated with the N radio-lobe is on the far side of the galaxy disk.
Therefore the S radio-lobe has $\phi~>~0{\arcdeg}$.

\vspace{-0.15in}

\section{Discussion}

We have shown that the distributions of $\beta$ for
Seyfert~1's, Seyfert~2.0's, and all Seyferts taken together are consistent
with the hypothesis that the radio jets are randomly oriented with respect 
to the galaxy disk (Fig.~8). This hypothesis is also supported by the 
distribution of $\beta$ for Seyfert~1's and Seyfert~2.0's, under the 
assumptions of the unified scheme (Fig.~9).
However, the distribution of $\delta$ (Figs.~3 and 4) for
Seyfert~2.0's may be different from a uniform distribution, but only at the
$\simeq$90\% confidence level (Table~4). This possible non-uniformity shows
up as an apparent deficiency of Seyfert~2.0's with 
70{\arcdeg}~$<~\delta~<~$90{\arcdeg}, as already noted by previous workers
(Paper VI; S97). 
We now briefly discuss possible causes of this deficiency.

\vspace{-0.15in}

\subsection{The Possible Deficiency of Seyfert~2.0's at Large $\delta$}
In order to determine which jets contribute to the possible deficiency of
Seyfert~2.0's at large $\delta$, we have computed the distribution of galaxies
as a function of $\beta$ and $\theta$ for galaxies  with
$\delta~\geq~70{\arcdeg}$. We have used $\phi_c$~= 30{\arcdeg} and the 
observed distribution of $i$ (Figure~7). The resulting distribution, shown
in Figure~10, is asymmetric about the $\theta$~= 0{\arcdeg} plane because
most of the jets near $\theta$~= $-$90{\arcdeg}, $\beta$~= 40{\arcdeg}
are directed towards the observer and therefore classified as Seyfert~1's. 
The central ``valley floor'' at the smaller values of $\mid\theta\mid$
and $\beta$~$\geq$ 20{\arcdeg} appears because these Seyfert~2.0's have 
$\delta$~$\leq$ 70{\arcdeg}.
Varying the value of $\phi_c$ and the weight distribution of $i$
does not significantly change the results that  : \newline
(i)~all values of $\beta$ contribute to the possible deficiency of 
Seyfert~2.0's at high $\delta$, so this deficiency cannot result from a 
simple absence or excess of galaxies with certain values of $\beta$; \newline
(ii)~an absence of jets with 
70{\arcdeg}~$\lesssim~\mid\theta\mid~\lesssim$~90{\arcdeg} 
at all $\beta$ values would produce the observed deficiency, but since the 
angle $\theta$ is defined by the observer's orientation, there is no physical
basis for such an absence. \newline

\begin{figure}[!ht]
\figurenum{10}
\plotfiddle{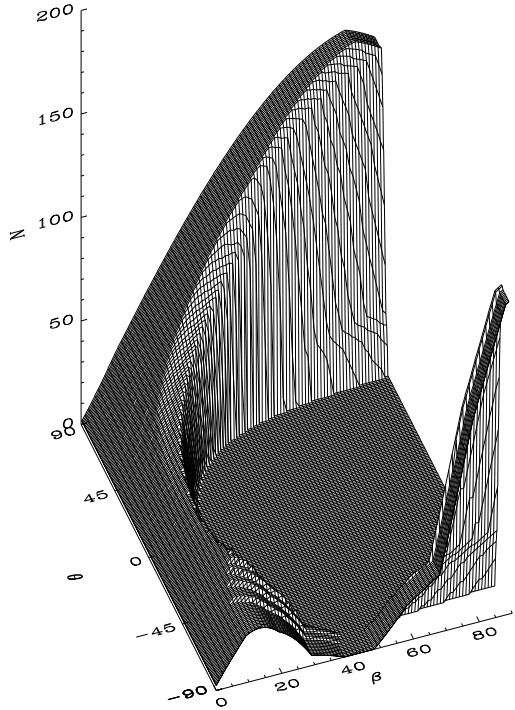}{4in}{0}{50}{50}{-150}{-40}
\caption
{\small{
The calculated distribution as a function of $\beta$ 
and $\theta$ for Seyfert~2.0 galaxies with $\delta~\geq$ 70{\arcdeg}.
We have used $\phi_c$~=~30{\arcdeg}, and the observed distribution of
Seyfert~2.0 inclinations. 
}}
\end{figure}

The best explanation for the observed distribution of $\delta$ is a 
smooth decline in the number of galaxies from $\delta$~= 0{\arcdeg} to
$\delta$~= 90{\arcdeg}, as suggested earlier by S97.
Such a $\delta$ distribution can result from a $\beta$ distribution which 
has relatively few jets at $\beta~\leq~30{\arcdeg}$ and/or relatively more
jets at $\beta~\geq~60{\arcdeg}$ (see Figure~6c of S97).
We have computed the expected distribution of $\delta$ by starting with a
random distribution of jets and then taking 50\% fewer jets at
$\beta~\leq~30{\arcdeg}$ and 50\% more jets at $\beta~\geq~60{\arcdeg}$, 
a distribution consistent with the points for Seyfert 2.0's in Figures~8 and 9.
We use $\phi_c$~= 30{\arcdeg} and the observed distribution of $i$ (Fig.~7).
The resulting expected  distributions of $\delta$ are shown with the thick
dashed lines in the two left panels
of Figure~4, and are consistent with the observations.

A distribution which favors larger values of $\beta$ can result from
any of the following effects:
(i)~an asymmetric distribution of ambient density in the central kpc, as
mentioned by S97. If the density were lower along the disk axis, plasma 
ejecta with low $\beta$ might produce less radio emission, if interaction
with the ambient medium is needed for significant radio synchrotron emission.
(ii)~contamination by SNR's and other emission in the disk of the galaxy, 
which would be interpreted as radio ejecta at $\beta$~=~90{\arcdeg}.
(iii)~gravitational bending of low velocity ,`heavy', radio ejecta by the
gravitational potential of the disk (``fountain model''). For example, a 
plasmoid moving ballistically with an ejection velocity of 500~km~s$^{-1}$ 
at $\beta$~=~45{\arcdeg} in a disk potential of 
$6.5~\times~10^8$ M$_{\sun}$~kpc$^{-2}$ will be bent to $\beta$~=~64{\arcdeg}
in 2~$\times~10^7$~yr. This effect will change an initially random jet
distribution to one which favors high values of $\beta$. 
(iv)~a preferential orientation of galactic magnetic field perpendicular
to the disk over the inner $\sim$kpc of the galaxy, 
which would cause radio ejecta at low $\beta$ values to see a smaller
component of tangential magnetic field. Models which use ambient field
compression (e.g. \cite{wilulv87}) predict lower synchrotron emission from 
such ejecta.
(v)~a trend for the accretion disks to be preferentially highly inclined 
with respect to the galaxy stellar disks. While a process such as 
the radiatively-driven disk instability (\cite{pri97}) may serve to
{\it{randomize}} the planes
of the nuclear accretion disks, we know of no process which would lead
to the accretion and galaxy stellar disks systematically {\it{avoiding}}
being coplanar.

\subsection{The Unified Scheme}
Our analysis shows that the distributions of $\beta$ for Seyfert~1 and 2.0 
galaxies are consistent with the unified scheme. There is, however, no 
evidence which favors the unified scheme over any alternative explanation.
If the distribution of $\beta$ in a larger sample
of Seyfert~1's is found to significantly
disfavor higher values of $\beta$, then this, 
in light of the absence of highly inclined Seyfert~1's, 
will provide strong support to the idea that the axes of the accretion disks
in Seyfert~1's are preferentially oriented along the line of sight. 

\section{Conclusions}
We have used the distribution of $\delta$ and $i$ 
to investigate the distribution of $\beta$ in two samples,
the early-type Seyfert sample and the sample of all radio-extended
Seyferts. Our ability to derive the distribution of $\beta$ is 
limited by observational uncertainties and biases, and several
selection effects within the Seyfert samples.
An analysis of the $\delta$ distributions shows that the distribution
for Seyfert~2.0's is only marginally different from uniform (at the $\sim$90\%
confidence level), while that of Seyfert~1's is not significantly different 
from uniform. The $\delta$ distributions of Seyfert~1's and 2.0's are not
significantly different from each other, though all results are limited by
the small number of objects, especially Seyfert~1's.

The distribution of $\delta$ for non-interacting late-type (Sab and later) 
Seyferts may favor large values of $\delta$ (at the $\sim$96\% 
confidence level), while that for early-type Seyferts is found to be more 
or less random. We suggest that this result indicates that  
the nuclear accretion disk in non-interacting late-type spirals
shares the angular momentum direction of the stellar disk, while,
in early-type galaxies,
the accretion disk is randomly oriented with respect to the
stellar disk and may have resulted from accretion following a galaxy merger.

We followed the method of C98 to find that the distributions of $\beta$ derived 
from all Seyfert data together, and from the Seyfert~1 and 2.0 data separately,
are consistent with the hypothesis that the jets are randomly oriented.
Following C98, we also used the unified scheme to calculate the distributions
of $\beta$ for Seyfert~1's and 2.0's separately. The resulting distributions
are different, but this difference may be ascribed to the absence of highly
inclined galaxies in Seyfert samples, especially type 1 Seyferts. 
There is, therefore, no evidence that the $\beta$ distributions of Seyfert~1's
and 2.0's are different from each other or that either one is significantly
different from a random distribution. The results are also consistent with 
the unified scheme.

Finally, we discussed the possible non-uniformity of the $\delta$ distribution
for Seyfert~2.0's.  The most straightforward interpretation is that the
distribution of $\delta$ for Seyfert~2.0's decreases smoothly from
$\delta$~=~0{\arcdeg} to $\delta$~=~90{\arcdeg}. 
Such a distribution can occur if larger values of $\beta$ are somewhat more 
common than smaller values, and various possible physical explanations were
discussed.
A much larger sample will be required for a more definitive evaluation of
this issue.

\acknowledgements
NN would like to thank Pierre Ferruit and Carole Mundell for helpful 
discussions,  and Jim Ulvestad for kindly supplying unpublished radio maps.
We thank the referee, Ski Antonucci, for detailed comments which helped
to significantly improve the manuscript.
This research has made use of the NASA/IPAC extragalactic database (NED) which
is operated by the Jet Propulsion Laboratory, Caltech, under contract
with the National Aeronautics and Space Administration.
We have made use of the Lyon-Meudon Extragalactic Database                   
(LEDA) supplied by the LEDA team at the CRAL-Observatoire de                   
Lyon (France).           
We have used the Digital Sky Surveys (DSS) which 
were produced at the Space Telescope Science Institute under U.S. 
Government grant NAG W-2166.
The images of these surveys are based on photographic data obtained
using the Oschin Schmidt Telescope on Palomar Mountain and the UK Schmidt
Telescope. The plates were processed into the present
compressed digital form with the permission of these institutions. 
This research has made use of the statistical tests in ASURV 1.2
(\cite{lif92}).
This work was supported by grant AST~9527289 from the National Science
Foundation and grant NAG~81027 from NASA.

\newpage
\appendix
\section{Notes on Seyfert Galaxies not in Early-Type Sample}

\paragraph{NGC~1068}
Radio source S1 is generally believed to represent the location of the
active nucleus, based on its spectrum (\cite{galet96}),
morphology (Gallimore, Baum \& O'Dea 1997) and the presence of water vapor
masers (\cite{greet96}). 
The P.A. between sources S1 and S2 (Gallimore et al. 1996; their Figure~1)
is $-$6{\arcdeg} while the P.A. between sources S1 and C is 13{\arcdeg}.
We adopt a radio P.A. of 0{\arcdeg}.  RC3 lists a photometric major axis
P.A.$_{RC3}$~= 70{\arcdeg} and log~R$_{25}$~= 0.07. 
HI kinematic data (\cite{brimun96}) show that the galaxy disk is warped, 
with a major axis P.A. (P.A.$_{HI}$) that increases from 95{\arcdeg} in the
high surface brightness inner ring
to 115{\arcdeg} in the faint outer extensions at radius $\sim$200{\arcsec}.
Thus, P.A.$_{RC3}$ - P.A.$_{HI}$~= 45{\arcdeg}.

\addtolength{\parskip}{-0.25in}

\paragraph{NGC~1144} (Arp 118)
This galaxy shares a common envelope with NGC~1143.
P.A.$_{RC3}$~= 110{\arcdeg} for NGC~1143 and P.A.$_{RC3}$~= 130{\arcdeg} for 
NGC~1144.

\paragraph{NGC~1365} RC3 lists major axis P.A.$_{RC3}$~= 32{\arcdeg} and 
log~R$_{25}$~= 0.26.
The HI kinematic studies of Ondrechen \& van der Hulst (1989) indicate a 
major axis P.A.$_{HI}$ of 222{\arcdeg} and an inclination of 46{\arcdeg}.
Thus, P.A.$_{RC3}$ - P.A.$_{HI}$~= 10{\arcdeg}.

\paragraph{NGC~2622} 
RC3 lists  
log~R$_{25}$~= 0.24 but does not give a major axis P.A.
A second generation DSS image clearly shows that the galaxy is interacting
with its nearest neighbor which is at a distance of 1{\arcmin}.

\paragraph{NGC~2992} (Arp 245, MCG$-$2-25-14).
This galaxy forms an interacting pair with NGC~2993. 
RC3 lists log~R$_{25}$~= 0.51, but does not give a major axis P.A.
The ``Extended Southern Galactic Catalog'' 
(Corwin et al. 1998, hereafter ESGC), 
lists galaxy major axis P.A.~= 15{\arcdeg} and 
diameter~= 6{\farcm}03~x~2{\farcm}19 in B. 
Thompson \& Martin (1988) used enlarged Sky Survey prints to measure a
major axis P.A. of 18{\arcdeg}.
DSS images show that while the galaxy is strongly interacting,
the central and northern parts
of the galaxy appear undisturbed and the isophotes show a consistent 
major axis P.A. 

\paragraph{NGC~3227} (Arp 94).  RC3 lists major axis P.A.$_{RC3}$~=
155{\arcdeg} and  log~R$_{25}$~= 0.17. 
We use the major axis P.A. and inclination obtained
from HI kinematic observations (\cite{munet95}).
P.A.$_{RC3}$ - P.A.$_{HI}$~= 3{\arcdeg}.

\paragraph{NGC~5135} (ESO 444-G32, MCG$-$5-32-13)
RC3 lists log~R$_{25}$~= 0.15 but does not list a major axis P.A.
ESO lists a diameter of 3{\farcm}5~x~3{\farcm}3.
Corwin et al. (1985) find the galaxy diameter to be
5{\farcm}5~x~5{\farcm}37. 
A first generation DSS image confirms that the galaxy is nearly
circular.

\paragraph{NGC~5643} 
RC3 lists log~R$_{25}$~= 0.06 and does not list a major axis P.A.
The ESO catalog lists the galaxy diameter as 6{\arcmin}~x~5{\farcm}5 in B. 
de Vaucouleurs (1977) gives a galaxy diameter of 
5{\farcm}37 x 4{\farcm}79, while
Corwin et al. (1985) give 6{\farcm}92 x 6{\farcm}46.
Morris et al. (1985) measure a photometric major axis P.A. of
128{\arcdeg}$\pm$10{\arcdeg} and a  kinematic major axis P.A. of 
136{\fdg}5$\pm$2{\fdg}5.

\paragraph{NGC~5728}
RC3 lists major axis P.A.$_{RC3}$~= 30{\arcdeg}, which is the P.A. of the 
bar.
ESGC lists major axis P.A.~= 0{\arcdeg} at a diameter of
6{\farcm}92 x 5{\farcm}37 in B.  
DSS images show that the major axis P.A. is close to 0{\arcdeg}.
Schommer et al. (1988) find the galaxy to have a 
photometric major axis P.A. of 2{\arcdeg}$\pm$5{\arcdeg} and a
kinematic major axis P.A. of $\sim$2{\arcdeg}.

\paragraph{NGC~5929} 
This galaxy shares a common outer envelope with its
neighbor NGC~5930.
RC3 lists log~R$_{25}$~= 0.03 and does not list a major axis P.A.
UGC lists a galaxy diameter of 1{\farcm}1~x~1{\arcmin} in R and
1{\arcmin}~x~0{\farcm}9 in B. 
The galaxy is too round to measure a major axis P.A.

\paragraph{NGC~6814}
RC3 does not list a major axis P.A. for this object. 
ESGC lists a diameter of 4{\farcm}47~x~4{\farcm}47
and Buta (1988, private communication to ``The Lyon-Meudon Extragalactic
Database'' see e.g. Paturel et al. 1997, hereafter LEDA)
finds  a diameter of 3{\farcm}02~x~2{\farcm}88.
Liszt \& Dickey (1995) have mapped the HI emission in this galaxy
out to a diameter of about 6{\arcmin}. They find a large kinematic warp 
in the outer galaxy disk: the kinematic
major axis decreases monotonically from 200{\arcdeg} at distance 3~kpc
from the center to 165{\arcdeg} at distance 21 kpc from the
center, leading them to adopt a major axis P.A.$_{HI}$~= 176{\arcdeg}. 
They also suggest that the galaxy is much more nearly face-on than
$i$~=~22{\arcdeg}, and we therefore label its inclination as 
$i$~=~0{\arcdeg} (see Section 2.4).

\paragraph{NGC~7450} (Mrk~1126) 
RC3 lists log~R$_{25}$~= 0 while 
ESGC lists a galaxy diameter of 2{\farcm}19~x~2{\farcm}19
in B. 
A second generation DSS image confirms that the galaxy is almost circular.

\paragraph{Mrk~34} 
This galaxy is not listed in the RC3 and UGC catalogs.
Garnier et al. (1996) find the galaxy to have a diameter of 
0{\farcm}63~x~0{\farcm}44 in B. 
A second generation DSS image shows the galaxy is compact, with
ill-defined outer isophotes.

\paragraph{Mrk~78}
RC3 lists log~R$_{25}$~= 0.26, but does not give a galaxy major axis P.A.
Takase \& Miyauchi-Isobe (1987) list a galaxy diameter of 
0{\farcm}4~x~0{\farcm}2.
A first generation DSS image shows the outer isophotes to be ill-defined.

\paragraph{Mrk~79} 
Oke and Lauer (1979) state that this galaxy is extended along 
P.A.$\simeq$65{\arcdeg}.
MacKenty (1990) measures a major axis P.A. of 65{\arcdeg} at the
24 mag (arcsec)$^{-2}$ contour in R corresponding to a major axis diameter of 
0{\farcm}8.
The image published in MacKenty (1990)
shows an anomalous arm which starts to
the NE and then curves to the SE. This is the most extended feature
in the image and biases any attempt to determine a major axis P.A.
UGC lists galaxy diameters of 1{\farcm}41~x~1{\farcm}41 in B and 
1{\farcm}51~x~1{\farcm}29 in R  and 
Garnier et al. (1996) measure a diameter of 1{\farcm}07~x~1{\farcm}07
in B. 
Since these measurements are at a larger diameter than those in MacKenty
(1990), we consider the galaxy to be essentially circular.

\paragraph{Mrk~110}
This galaxy is not listed in the RC3 and UGC catalogs.
Takase \& Miyauchi-Isobe (1987) list a galaxy diameter of 
0{\farcm}5~x~0{\farcm}3.
while MacKenty (1990) finds this object is circular
at the 24~mag~(arcsec)$^{-2}$ isophote in R, which corresponds to
a diameter of 24{\arcsec}.
A second generation DSS image shows that the outer isophotes of the
galaxy are disturbed, with a prominent tail to the west.

\paragraph{Mrk~176} (Arp 322)
RC3 lists log~R$_{25}$~= 0.53 but does not give a major axis P.A.
Garnier et al. (1996) measure a major axis P.A. of 59{\arcdeg} and a galaxy 
extent of 0{\farcm}81~x~0{\farcm}41 in B. 
A second generation DSS image shows that Mrk~176 shares a common
outer envelope with two close companion galaxies.

\paragraph{Mrk~231}
The radio morphology of this source is complex. Recent VLBA
images (Ulvestad, Wrobel \& Carilli 1998) reveal a 2~pc scale
`jet' in P.A. 65{\arcdeg}, and a 40~pc scale triple radio structure 
in P.A. 0{\arcdeg} which is the P.A. of the larger scale
VLBA and VLA radio structure.
There is no evidence of radio structure linking the 2~pc and 40~pc scale
radio knots.
We use a radio P.A. of 0{\arcdeg} in order to be consistent with the 
scale over which the radio P.A. has been measured for most other sources in
the sample.

\paragraph{Mrk~266} (NGC~5256) 
This galactic pair consists of a Seyfert~2 nucleus to the SW  and
a LINER nucleus to the NE.
RC3 lists log~R$_{25}$~= 0.05 but does not give a major axis P.A.
UGC lists an extent of  1{\farcm}2~x~1{\farcm}1
in B and Garnier et al. (1996) find a diameter of 
1{\farcm}17~x~1{\farcm}17 in B.
The galaxy is thus essentially circular.
Kukula et al. (1995) give a P.A. of 32{\arcdeg} for
the overall radio structure between the Seyfert and LINER
nucleus. Mazzarella et al. (1988) have used 
deeper high resolution VLA maps to find the radio structure 
associated with the Seyfert nucleus has an `L' type morphology in P.A.
$-$10{\arcdeg}.

\paragraph{Mrk~270} 
Both RC3 and UGC indicate this galaxy is nearly circular. We have
confirmed this with a second generation DSS image.

\paragraph{Mrk~273}
RC3 lists log~R$_{25}$~= 0.64 but does not give a major axis 
P.A. (the galaxy is classified as peculiar).
UGC lists a diameter of 1{\farcm}2~x~0{\farcm}25 in B but also
does not list a major axis P.A.
Garnier et al. (1996) give a major axis P.A.
of 11{\arcdeg} at a diameter of 1{\farcm}62~x~0{\farcm}52. 
A second generation DSS image shows that the most
prominent feature is a peculiar extension to the S.

\paragraph{Mrk~423}
This galaxy is not listed in UGC or ESO. RC3 lists log~R$_{25}$~= 0.23 
but does not give a major axis P.A.
DSS images show  an extension in P.A. 58{\arcdeg} at an extent of
about 1{\arcmin}~x~0{\farcm}5. The extension appears to be 
a pair of spiral arms and the image is not deep enough to 
determine a major axis P.A. for the putative disk.
 
\paragraph{Mrk~463E} 
The other component of the pair, Mrk~463W, is either a 
Seyfert~2 or a starburst galaxy (\cite{met91}). 
Neff \& Ulvestad (1988) find the radio P.A.~= 10{\arcdeg}
for Mrk~463E, while Mrk~463W remains unresolved.
Mazzarella et al. (1991) have found a radio component 1{\farcs}3 S of
Mkn~463E, corresponding to a linear distance of 1.9 kpc.
RC3 does not give a major axis P.A. but lists log~R$_{25}$~= 0.34.
Thompson \& Martin (1988) used enlarged Sky Survey prints to measure a major
axis P.A. of 89{\arcdeg}. 
A second generation DSS image shows that while the inner isophotes are
elliptical, the outer isophotes show an extension to the
SW. This faint extension is the most extended feature in the galaxy and 
biases any measurement of the major axis.

\paragraph{Mrk~509} 
This galaxy is not listed in RC3 or UGC.
Thompson \& Martin (1988) used enlarged Sky Survey prints to measure a major
axis P.A. of 70{\arcdeg} at an extent of 0{\farcm}17~x~0{\farcm}1.
Phillips, Baldwin \& Atwood (1983) used deeper images to
measure a P.A. of  70{\arcdeg}$\pm$5{\arcdeg}.

\paragraph{Mrk~530}
RC3 lists the major axis P.A.$_{RC3}$~= 165{\arcdeg} and log~R$_{25}$~= 0.18.
UGC also gives the major axis P.A.~= 165{\arcdeg} at a diameter of 
1{\farcm}58~x~0{\farcm}89 in B.
ESGC lists the major axis P.A.~= 5{\arcdeg} 
at a diameter of 2{\farcm}57~x~1{\farcm}95.

\paragraph{Mrk~533} (NGC~7674) 
RC3 lists log~R$_{25}$~= 0.04 and does not give a major axis P.A. 
Paturel \& Petit (private communication to LEDA) derive a major axis P.A.
 of 150{\arcdeg}, which appears
to be the P.A. of the brighter isophotes. The fainter isophotes are more
circular and are affected by a nearby companion to the NE.

\paragraph{Mrk~609}
RC3 lists  log~R$_{25}$~= 0.16, but does not give a major axis P.A.
ESGC lists major axis P.A.~= 90{\arcdeg} at a
diameter of 0{\farcm}91~x~0{\farcm}6 in B.
A first generation DSS image shows the major axis of the outer isophotes 
is ill defined.

\paragraph{Mrk~618}
RC3 lists log~R$_{25}$~= 0.12 but does not give a major axis P.A.
ESGC lists major axis P.A.~= 85{\arcdeg} at a
diameter of 1{\farcm}32~x~1{\farcm}02 in B.
A second generation DSS image shows the galaxy to be nearly circular.

\paragraph{Mrk~926}
This galaxy is not listed in the RC3, UGC and ESO catalogs.
Garnier et al. (1996) measure a diameter of 0{\farcm}78~x~0{\farcm}54 in V. 
We used a first generation DSS image to measure a major axis 
P.A. of 90{\arcdeg} at a diameter of 0{\farcm}71~x~0{\farcm}64.

\paragraph{MCG 8-11-11}
RC3 and UGC list major axis P.A.~= 90{\arcdeg}. 
UGC lists a diameter of 2{\farcm}82~x~2{\farcm}51 in B.
A second generation DSS image
shows the galaxy to be near circular at a diameter of about 2{\arcmin}.

\paragraph{ESO428-G14} 
This is listed as 0714-2914 in Paper VII.

\paragraph{UGC 5101}
UGC lists major axis P.A.~= 87{\arcdeg} and an extent of 
1{\farcm}2~x~0{\farcm}6 in B and 
1{\farcm}2~x~0{\farcm}79 in R.
A first generation DSS image reveals that while the brighter isophotes
of the galaxy are elliptical, part of the emission along the major axis 
is made up of a narrow $\sim$30{\arcsec} finger extending to the west.

\newpage

\addtolength{\parskip}{-0.15in}

\newpage

\pagestyle{empty}

\begin{figure}
\plotfiddle{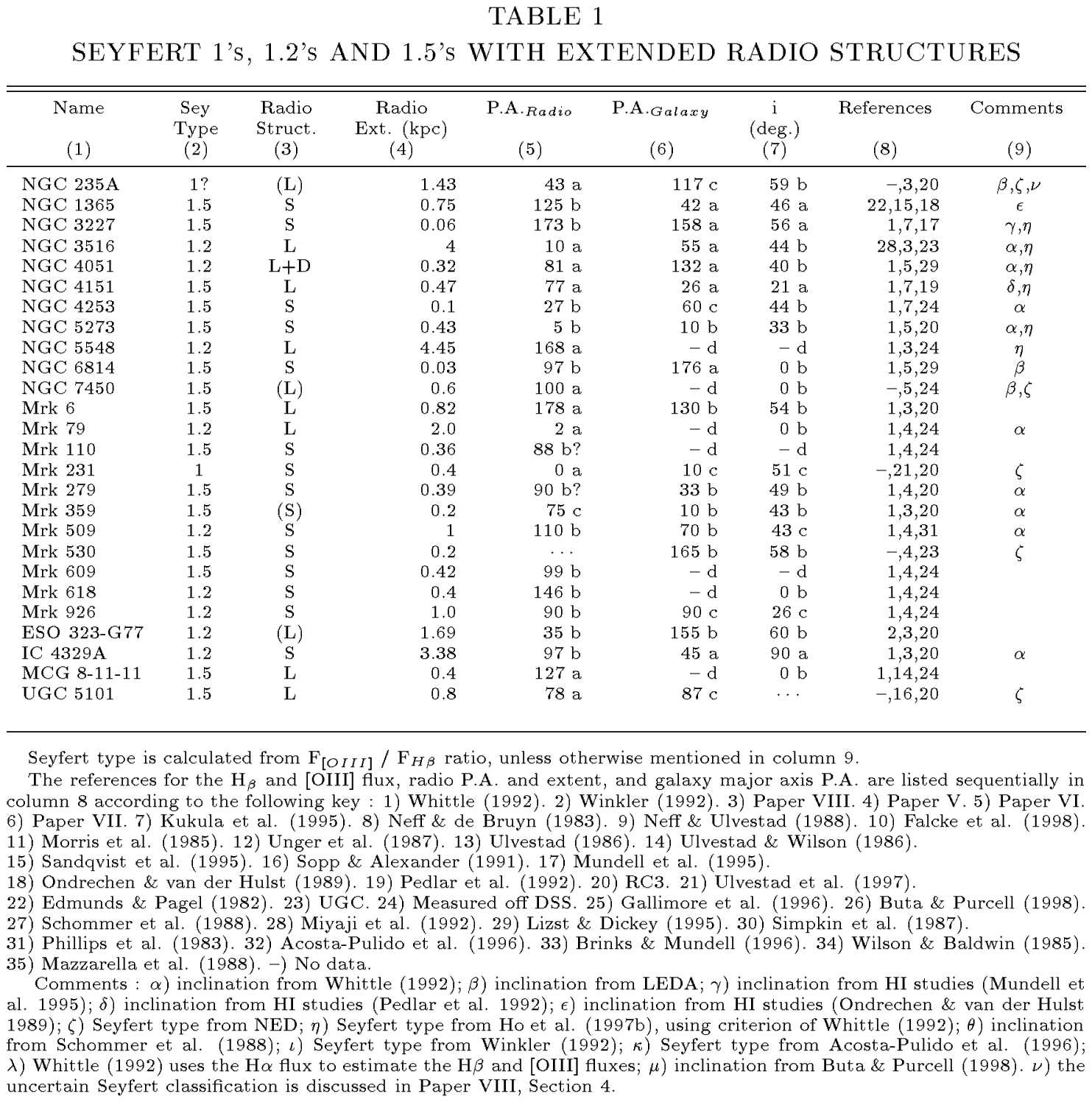}{9in}{0}{100}{100}{-300}{70}
\plotfiddle{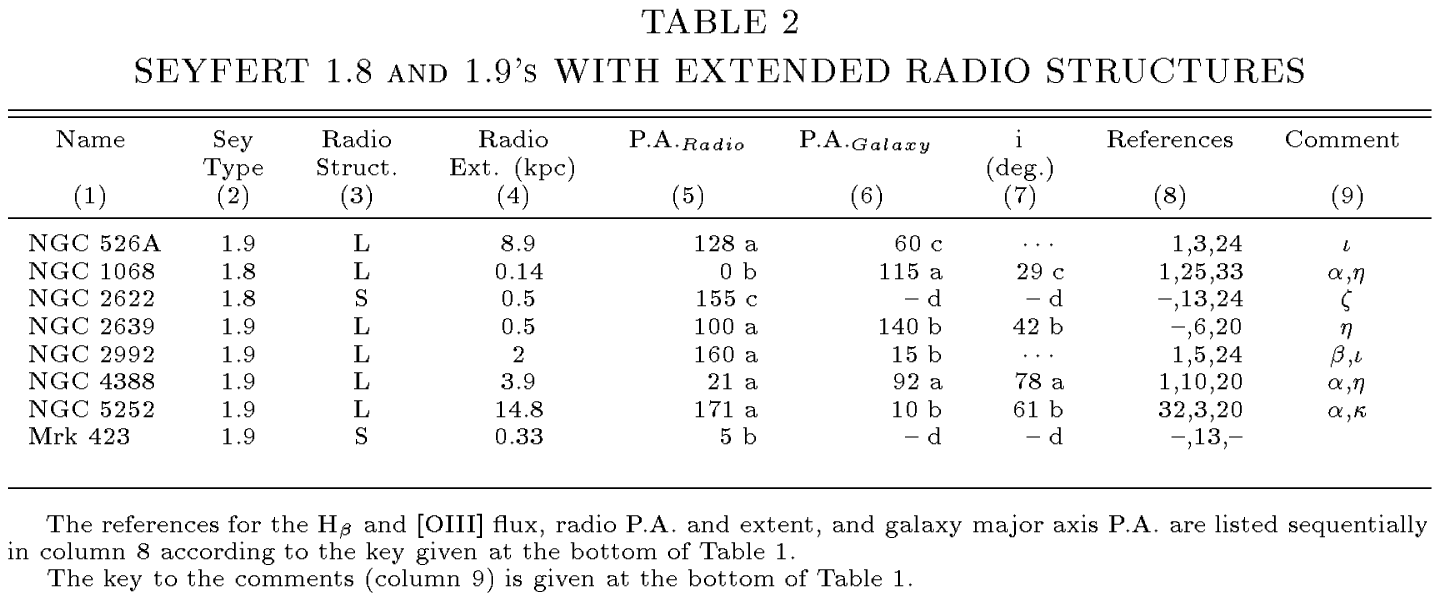}{9in}{0}{100}{100}{-300}{400}
\end{figure}

\newpage

\begin{figure}
\plotfiddle{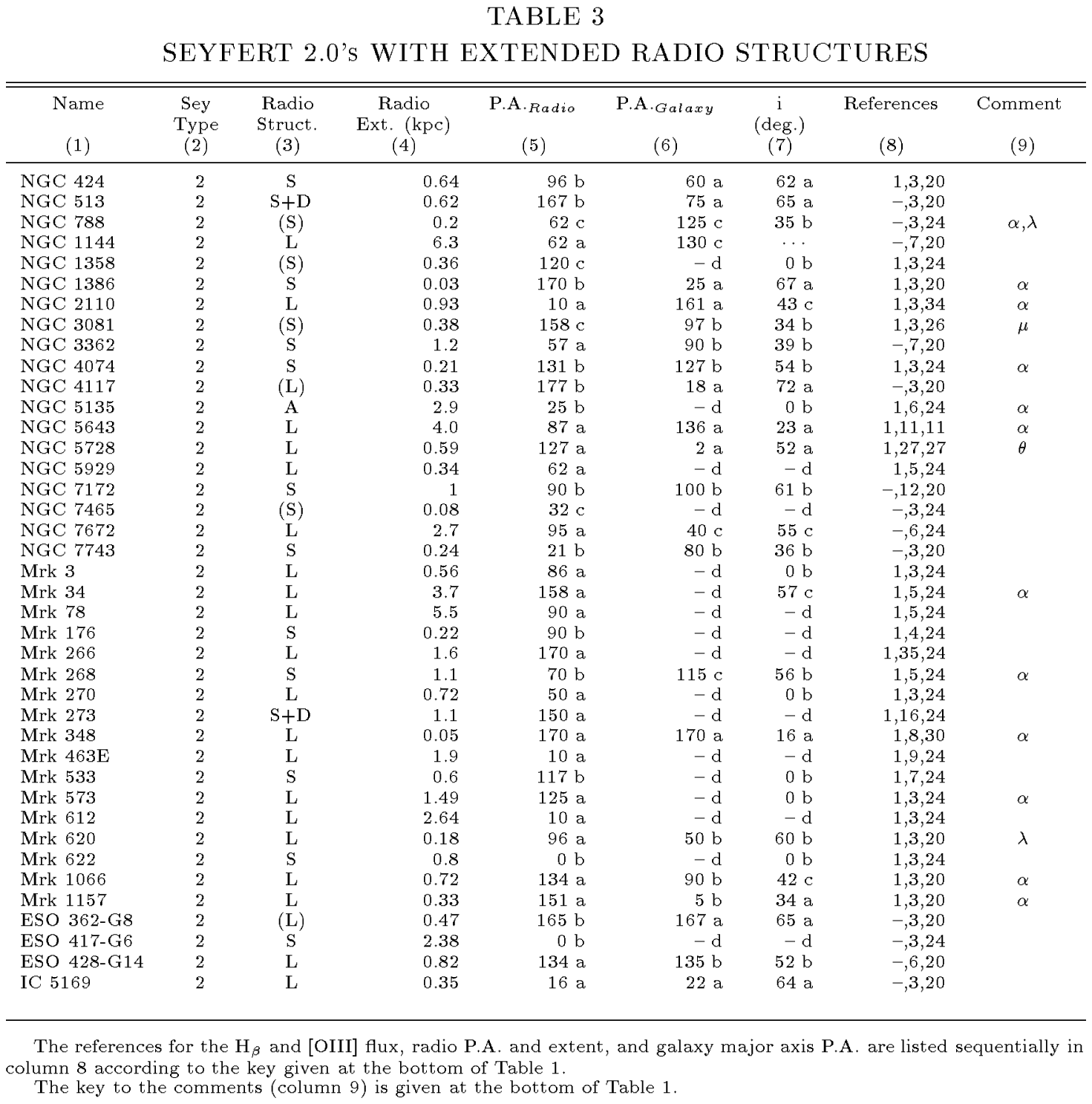}{9in}{0}{100}{100}{-300}{0}
\end{figure}

\newpage

\begin{figure}
\plotfiddle{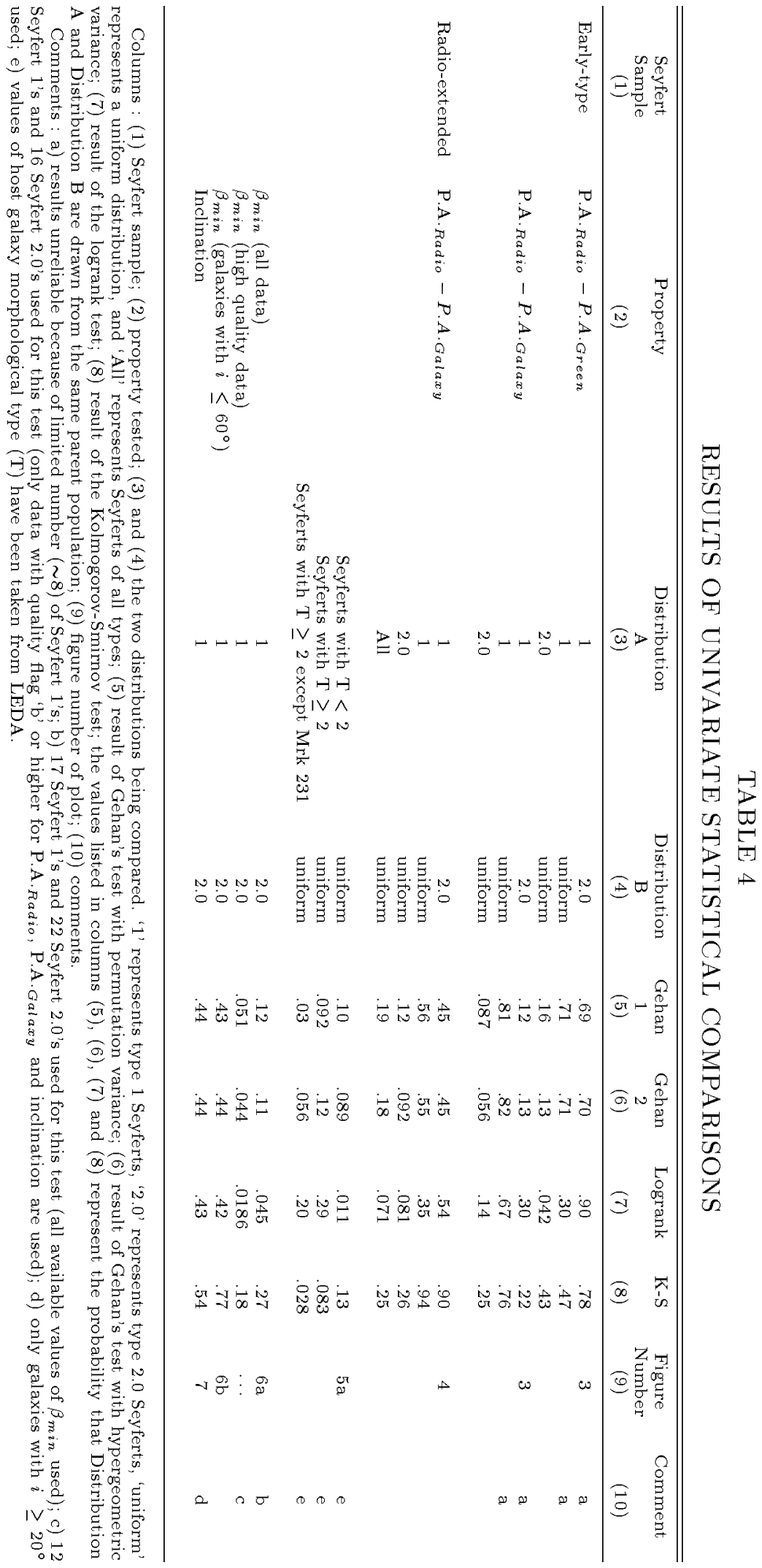}{9in}{180}{92}{92}{200}{730}
\end{figure}

\end{document}